\documentclass[reprint,superscriptaddress,nofootinbib,amsmath,amssymb, aps,onecolumn]{revtex4-1}

\usepackage[a4paper, margin=2.50cm]{geometry}

\usepackage{graphicx}
\usepackage{fancyhdr, color}
\usepackage{hyperref} 

\usepackage[T1]{fontenc}
\usepackage{graphicx}
\usepackage{dcolumn}
\usepackage{bm}
\usepackage{subfig}
\usepackage{array}
\usepackage{framed}
\usepackage{verbatim}
\usepackage{graphicx}
\usepackage{esint}

\newcommand{\beq}{\begin{equation}}
\newcommand{\eeq}{\end{equation}}
\newcommand{\beqar}{\begin{eqnarray}}
\newcommand{\eeqar}{\end{eqnarray}}
\newcommand{\bea}{\begin{eqnarray}}
\newcommand{\eea}{\end{eqnarray}}
\newcommand{\bcen}{\begin{center}}
\newcommand{\ecen}{\end{center}}

\newcommand{\bra}[1]{\left< #1 \right|}
\newcommand{\ket}[1]{\left| #1 \right>}
\newcommand{\ketbra}[2]{\left| #1 \right> \left< #2 \right|}
\newcommand{\braket}[2]{\left< #1 \vert #2 \right>}

\newcommand{\ave}[1]{\left< #1 \right>}

\newcommand{\half}{\frac{1}{2}}

\newcommand{\Lc}{\mathcal{L}}
\newcommand{\Lch}{\mathcal{L}_h}
\newcommand{\Lcc}{\mathcal{L}_c}

\newcommand{\bs}{\bar{s}}

\newcommand{\mc}[1]{\mathcal{#1}}

\begin{document}

\fancyhead[R]{}

\title{Quantum features and signatures of quantum-thermal machines}

\author{Amikam Levy}
\email{amikamlevy@gmail.com} 

\affiliation{Department of  Chemistry, University of California Berkeley,  Berkeley, California 94720, USA.}

\affiliation{The Sackler Center for Computational Molecular Science, Tel Aviv University, Tel Aviv 69978, Israel.}

\author{David Gelbwaser-Klimovsky}
\email{dgelbi@yahoo.com.mx}

\affiliation{Department of Chemistry and Chemical Biology, Harvard University, Cambridge, MA 02138, USA.}

\date{\today}

\begin{abstract}

The aim of this book chapter is to indicate how quantum phenomena are  affecting the operation of microscopic thermal machines, such as engines and refrigerators.
As converting heat to work is one of the fundamental  concerns in thermodynamics, the platform of quantum-thermal machines sheds light on thermodynamics in the quantum regime. 
This chapter focuses on the basic features of quantum mechanics, such as energy quantization, the uncertainty principle, quantum coherence and correlations, and their manifestation in microscopic thermal devices. 
In addition to indicating the peculiar behaviors of thermal-machines due to their non-classical features, we present quantum-thermodynamic signatures of these machines. 
Any violation of the classical bounds on thermodynamic measurements of these machines  is a sufficient condition to conclude that quantum effects are present in the operation of that thermal machine. 
Experimental setups demonstrating some of the results are also presented.

\end{abstract}

\maketitle

\thispagestyle{fancy}

\section{Introduction}
Quantum mechanics is one of the greatest revolutions in the history
of science. 
It changed the way we understand the world and demonstrates
that the microscopic realm is governed by a theory that is fundamentally
different from classical mechanics. 
Classical mechanics is formulated on the phase space  and treats particles as points whereas quantum mechanics is described by wave functions and operators acting in  Hilbert space. 
In the quantum framework, the classical perception of certainty is replaced by probability, and instead of being continuous, physical quantities are generally  quantized.
These significant differences result in peculiar phenomena that are observed only in the quantum regime.
\par
Since the beginning, the development of quantum mechanics has been influenced by thermodynamics. 
 Planck initiated the quantum era by introducing the quantization hypothesis to describe thermal radiation emitted from a black body; Einstein discovered stimulated emission while studying  thermal equilibration between light and matter. 
In spite of the close historical relationship between quantum mechanics
and thermodynamics, one could wonder if the quantum revolution would eventually
shake the foundations of thermodynamics as it did with classical mechanics.
Some of the early works in this direction studied the behavior
of heat machines, which, besides their technological applications,
were used to test the compliance with the laws of thermodynamics.
Despite the effort focused on this direction, as of today, the fundamental laws and bounds of thermodynamics seem to hold also in the quantum regime. 
\par
Even though quantum mechanics complies with the   laws of thermodynamics, classical and quantum heat machines still diverge from one another in a non trivial way. 
As we show in this chapter, not only are classical and quantum heat machines fundamentally different, but quantum mechanics allows the realization of classically inconceivable heat machines. 
\par
In order to identify the appearance and effects of quantum features in thermal machines, one should first have a clear notion of classical thermal machines. Throughout this chapter, we will use different notions or levels of classicality, starting from the purest definitions, (i.e., a system is classical only if it is precisely described by classical mechanics), and working toward less strict definitions, which consider systems  governed by  quantum Hamiltonians as classical as long as either their coherences or quantum correlations are zero. 
In this manner we obtain a thorough understanding of how different quantum features influence thermal devices.   
\par
The chapter is divided into three sections concerning three main quantum features and their corresponding thermodynamic signatures. 
In the first section,  classicality will be considered as it is in classical thermodynamics, where the probability distribution is fully described by a phase space distribution without any constraint aside from normalization. In this section, quantum effects result only from  the quantization of energy levels and the uncertainty principle,  which sets some limitations on these probability distributions.
These features alone lead to discrepancies in the behavior of classical and quantum   heat machines.
\par
The second section describes the effects of quantum coherence in thermal machines.
Both the positive and negative implications of coherence are discussed. Furthermore,  we present a recent experiment demonstrating some of these results.
In these scenarios thermal machines that  are governed by a quantum Hamiltonian, but can be  fully described by their populations  are considered  classical (stochastic). 
In this context it's also important to note that the preferred basis to describe thermal machine is the one in which measurements are preformed, typically, this will be the energy basis.
\par
In the last section, we discuss the  role of correlations in the operation of thermal machines.  We demonstrate that quantum correlations can induce anomalous heat flow from a colder body to a hotter body, that can not be explained classically. 
Here, classicality refers to separable quantum states with zero discord, (i.e., only classical correlations are allowed). 
The chapter is structured so that each of the three main section is independent of the others.

\section{energy quantization and uncertainty principle}
\label{sec:energy_quant}

Classical and quantum mechanics provide different descriptions of
the same physical system. This is true even for the simplest cases.
As an example, consider a harmonic oscillator with mass $m$ and frequency
$\omega$, in a thermal state at temperature $T$. Classically, its
phase space distribution is given by the following Gaussian:
\begin{equation}
P_{HO}^{clas}(x,p)=\frac{\omega}{2\pi k_{B}T}e^{-\left(\frac{m\omega^{2}x^{2}}{2k_{B}T}+\frac{p^{2}}{2mk_{B}T}\right)}.
\label{eq:phoclas}
\end{equation}

Strictly speaking, phase space distributions are not part of the quantum
mechanics framework, as it is impossible to simultaneously determine
the position and momentum of a system. Nevertheless, quasi-probability
distributions, such as the Wigner function, share many properties
with phase distributions \cite{case2008wigner,wigner1932quantum} and are considered the ``closest
quantum equivalent'' to a phase space distribution. The Wigner function
for the same harmonic oscillator at the same state is
\begin{equation}
P_{HO}^{quan}(x,p)=\frac{\tanh[\frac{\hbar\omega}{2k_{B}T}]}{\hbar\pi}e^{-\tanh[\frac{\hbar\omega}{2k_{B}T}]\left(\frac{m\omega x^{2}}{\hbar}+\frac{p^{2}}{\hbar m\omega}\right)}.\label{eq:who}
\end{equation}

\begin{figure}[b]
\begin{centering}
\includegraphics[width=1\textwidth]{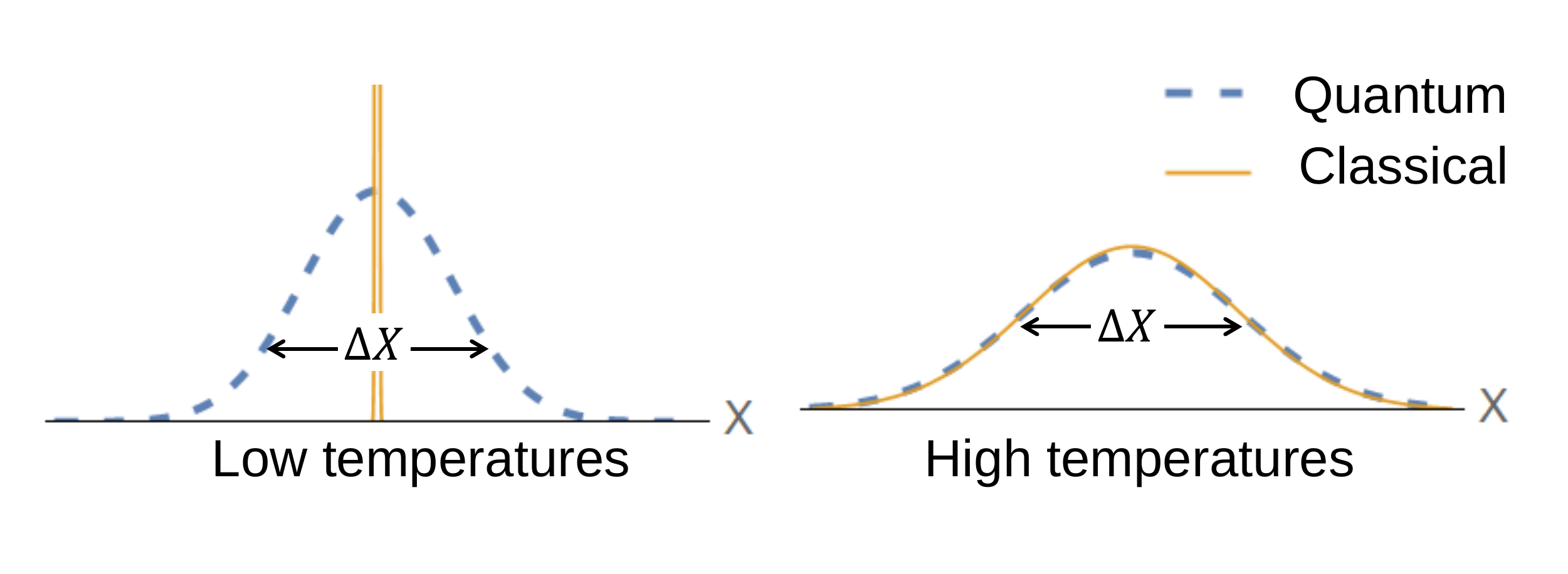} 
\par\end{centering}
\caption{Position projection of the classical  (yellow continuos line) and quantum  (blue dashed line) phase space distributions for a harmonic oscillator in a thermal state. Both distributions coincide at high temperatures (right), but at low temperatures (left) the quantum Wigner distribution, in contrast with the classical one, has to comply with the uncertainty principle, forcing the distributions to diverge.}
\label{fig:eneq:dist} 
\end{figure}

These two distributions coincide at the regime where the thermal energy
is much larger than the quantization energy, i.e., $\frac{\hbar\omega}{k_{B}T}\ll1.$
At this limit, the classical description is very precise, so quantum
effects can be neglected. Nevertheless, the distributions diverge
at low temperatures (see Fig. \ref{fig:eneq:dist}), where classical
mechanics predicts a smaller and smaller position and momentum uncertainties,
in contradiction with the uncertainty principle. In contrast, the
Wigner distribution, Eq. (\ref{eq:who}), obeys the uncertainty principle
at any temperature.

The differences between these distributions are based on the fact
that quantum systems have to comply with the uncertainty principle \cite{wigner1932quantum}.
Moreover, confining potentials impose boundary conditions that quantize
the energy state, create a zero point energy, and establish a dependence
of the system energy on the boundary conditions. The combination of
all these effects results on different energy, heat capacity and other
thermal properties \cite{sundar2017reproducing} relatively to the ones predicted by classical mechanics
and thermodynamics that neglect boundary effects \cite{callen1985thermodynamics}. In this section
of the chapter we will explore how those differences affect the performance
of thermal heat machines. We will compare exactly the same heat machine,
i.e., same working medium, same baths, same temperatures, etc., but
in one case the behavior of the working medium is dictated by classical
mechanics and in the other by quantum mechanics. In this section we
do not consider the effects of quantum coherences or quantum correlations.
These are left for subsequent sections of this chapter.

\begin{figure}
\begin{centering}
\includegraphics[width=1\textwidth]{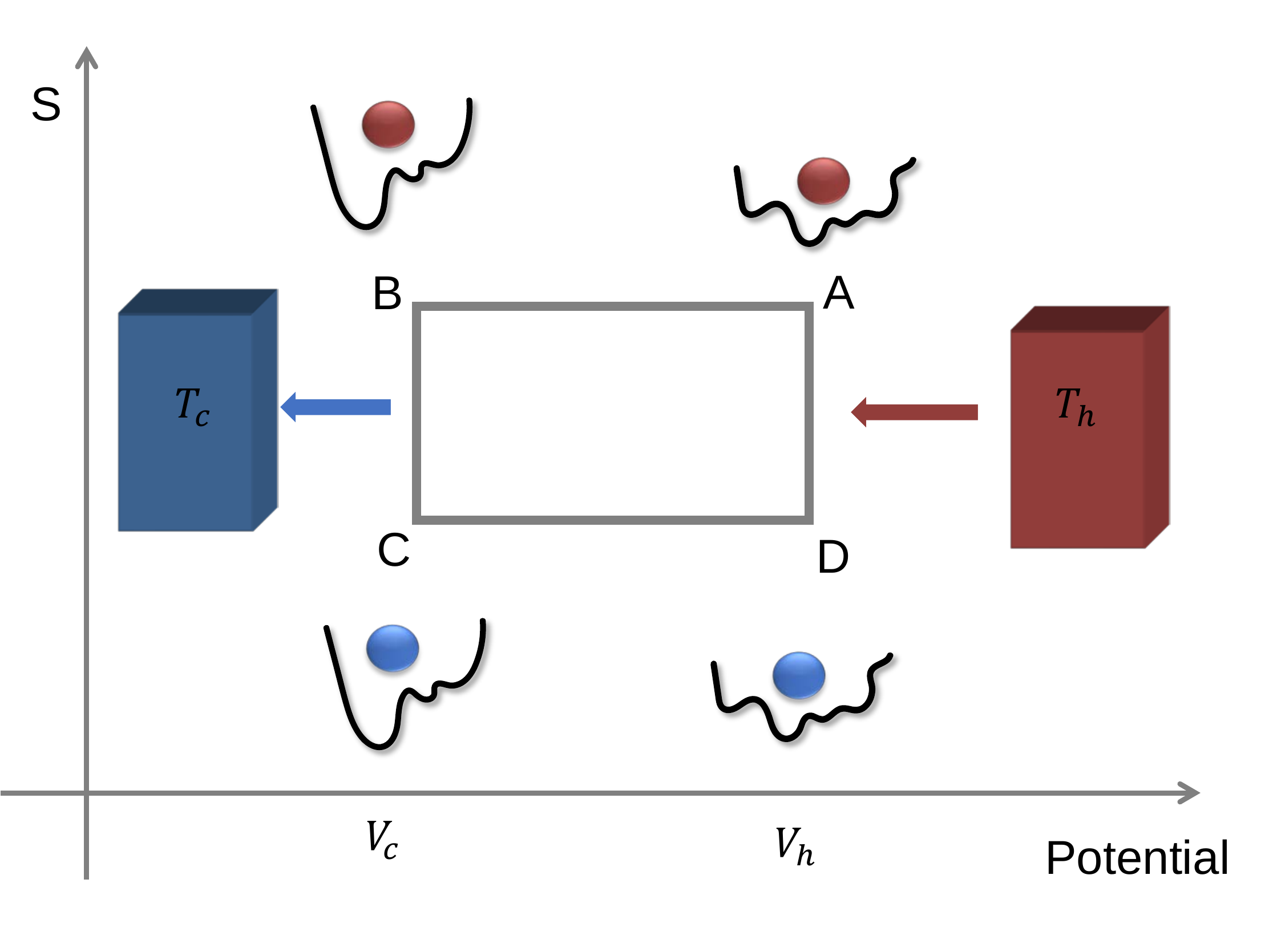} 
\par\end{centering}
\caption{Otto cycle. The x-axis represents a parameter that controls the working
medium potential and the y-axis represents its Von Neumann entropy.}
\label{fig:eneq:Otto} 
\end{figure}

In particular, we consider an Otto heat machine \cite{gelbwaser2017single,quan2007quantum,zheng2014work} operating with a M-dimensional
working medium, but a similar analysis can be done for other cycles.
The Otto cycle is composed of the following four strokes (see Fig.
\ref{fig:eneq:Otto}): At point A of the cycle the working medium
potential is $V_{h}(x_{1},...x_{M})\equiv V_{h}$. Its Hamiltonian
is $H_{h}$ and its energy levels are $\{E_{n}^{h}\}$. The working
medium is in thermal equilibrium with a hot bath at temperature $T_{h},$
so its state in the Hamiltonian energy basis is $\rho_{A}=Z_{T_{h}}^{-1}\sum_n e^{-\frac{E_{n}^{h}}{k_{B}T_{h}}}|n\rangle \langle n|,$
where $Z_{T_{h}}$ is the partition function. At $A$, the adiabatic
stroke starts: the working medium is decoupled from the hot bath and
the potential is slowly deformed until point B, where it reaches $V_{c}(x_{1},...x_{M})\equiv V_{c}$.
Here, the Hamiltonian is $H_{c}$, and its energy levels are $\{E_{n}^{c}\}$.
The deformation is slow enough for the process to fulfill the assumptions
of the quantum adiabatic theorem \cite{griffiths2016introduction}, so the levels populations
are the same $\rho_{B}=\rho_{A}$. At $B$, the system is coupled
to the cold thermal bath at temperature $T_{c}$, initiating the cold
isochoric stage. The system relaxes, achieving thermal equilibrium
at C, i.e., $\rho_{C}=Z_{T_{c}}^{-1}\sum_n e^{-\frac{E_{n}^{c}}{k_{B}T_{c}}}|n\rangle \langle n|,$
where $Z_{T_{c}}$ is the partition function. Then, the working medium
undergoes the second adiabatic stroke: it is decoupled from the bath
and the potential is again slowly transformed back, reaching $V_{h}$
at $D$. Here too, we apply the quantum adiabatic theorem so $\rho_{D}=\rho_{C}.$
Finally, at this point, the second isochoric phase starts: the working
medium is coupled again to the hot bath and equilibrates with it,
returning to point A.

One useful feature of the Otto cycle is that work and heat are exchanged
during different parts of the cycle: work is transferred during the
adiabatic processes and heat over the isochoric strokes. The heat flow
from the hot bath is
\begin{equation}
Q_{h}=\langle H_{h}\rangle_{A}-\langle H_{h}\rangle_{D}=\sum_{n}E_{n}^{h}\left(\frac{e^{-\frac{E_{n}^{h}}{k_{B}T_{h}}}}{Z_{T_{h}}}-\frac{e^{-\frac{E_{n}^{c}}{k_{B}T_{c}}}}{Z_{T_{c}}}\right)\label{eq:eneq:qh}
\end{equation}
and from the cold bath is
\begin{equation}
Q_{c}=\langle H_{c}\rangle_{C}-\langle H_{c}\rangle_{B}=\sum_{n}E_{n}^{c}\left(\frac{e^{-\frac{E_{n}^{c}}{k_{B}T_{c}}}}{Z_{T_{c}}}-\frac{e^{-\frac{E_{n}^{h}}{k_{B}T_{h}}}}{Z_{T_{h}}}\right).\label{eq:eneq:qc}
\end{equation}
Using the first law of thermodynamics, the expression for the work
can be obtained,
\begin{equation}
W=-Q_{h}-Q_{c}=\left(E_{n}^{c}-E_{n}^{h}\right)\left(\frac{e^{-\frac{E_{n}^{h}}{k_{B}T_{h}}}}{Z_{T_{h}}}-\frac{e^{-\frac{E_{n}^{c}}{k_{B}T_{c}}}}{Z_{T_{c}}}\right).\label{eq:eneq:w}
\end{equation}
Here, we use the sign convention that energy flowing to the working
medium is positive and flowing from the working medium is negative.
The heat machine has several operating modes: for $W<0$ and $Q_{h}>0$,
it is a heat engine and its thermal efficiency quantifies the amount
of work extracted per unit of incoming heat, i.e., $\eta=\frac{-W}{Q_{h}}$;
for $W>0$ and $Q_{c}>0$, it is a refrigerator and its efficiency
or coefficient of performance measures how much energy flows from
the cold bath per unit of invested work $COP=\frac{Q_{c}}{W}.$

For a classical Otto heat machine, whose working medium is an ideal
gas, the operation mode is determined by the compression ratio $r=\frac{Vol_{c}}{Vol_{h}}$,
($Vol_{c}$ is the container volume at point C and $Vol_{h}$ is the
container volume at point $A$): for $r>r_{Car}=\left(\frac{T_{h}}{T_{c}}\right)^{\frac{1}{\gamma-1}}$,
it operates as a refrigerator and for $1<r<r_{Car}$ the classical
Otto heat machine operates as an engine with an efficiency
\begin{equation}
\eta=1-\frac{1}{r^{\gamma-1}}\leq1-\frac{1}{r_{Car}^{\gamma-1}}=\eta_{Car},
\label{eq:efic}
\end{equation}
where $\eta_{Car}$ is the Carnot efficiency for an engine and $\gamma=\frac{C_{p}}{C_{v}}$
is the heat capacities ratio. An important property of the classical
Otto machine is that, in order to extract work or extract heat from
the cold bath, the working medium has to be compressible \cite{callen1985thermodynamics,farris1979rubber,mullen1975thermodynamics,paolucci2016continuum,gurtin2010mechanics}. Otherwise,
$r=1$ and $\eta=0$, precluding work extraction. This compression
ratio is smaller than $r_{Car}$, hindering also the refrigerator
operation. Here, quantum mechanics provides heat machines with an
advantage over their classical counterparts. As it was shown on \cite{gelbwaser2017single}
and we explain below, once the working medium is governed by quantum
laws instead of classical mechanics, compressibility is no longer
required for work extraction or for cooling down the cold bath. Quantum
mechanics allows the realization of heat machines with incompressible
working media, opening the possibility for creating classically inconceivable
heat machines.

One feature of a classical ideal gas undergoing an Otto cycle is
that it is always at equilibrium, so at every point of the cycle it
is possible to define the ideal gas temperature. This is not always
true for a quantum working medium. For example, at $B,$ the working
medium is at the state $\rho_{B}=Z_{T_{h}}^{-1}\sum_n e^{-\frac{E_{n}^{h}}{k_{B}T_{h}}}|n\rangle \langle n|,$
which is a Boltzmann distribution of the energy levels at point A
($\{E_{n}^{h}\}$) instead of point B ($\{E_{n}^{c}\}$). So, in general,
the state $\rho_{B}$ is not a thermal equilibrium state with respect to the Hamiltonian at point B. The only
exception is if the transformation homogeneously scales the energy
levels i.e., $E_{n}^{c}=qE_{n}^{h}$, where $q$ is independent of $n$. Under this condition we can
rewrite the state at B as $\rho_{B}=Z_{qT_{h}}^{-1}\sum_n e^{-\frac{E_{n}^{c}}{k_{B}qT_{h}}}|n\rangle \langle n|,$
which is a Boltzmann distribution at temperature $qT_{h}$ of the
energy levels at point B. In a similar way, the state at D, $\rho_{D}=Z_{T_{c}}^{-1}\sum_n e^{-\frac{E_{n}^{c}}{k_{B}T_{c}}}|n\rangle \langle n|=Z_{T_{c}/q}^{-1}\sum_n e^{-\frac{qE_{n}^{h}}{k_{B}T_{c}}}|n\rangle \langle n|$
is a Boltzmann distribution at temperature $T_{c}/q$ of the energy
levels at point D. Most of the research on quantum Otto cycles has
focused on this type of transformations \cite{quan2007quantum,zheng2014work} which includes
the length change of a 1D infinite well \cite{quan2009quantum}, frequency shift of a
1D harmonic oscillator \cite{kosloff17entropy} or any other scale invariant transformation \cite{zheng2015quantum}.
We will first study the differences between classical and quantum
heat machines undergoing homogeneously energy scaling, which allows
the working medium to be in a thermal state during the whole cycle.
Then, we will explore the regime of inhomogeneous energy scaling,
which has been seldom studied, with few exceptions \cite{uzdin2014multilevel,quan2005quantum}. For an inhomogeneous
energy scaling, the working medium deviates from a thermal state during
the adiabatic strokes, resulting on more striking discrepancies between
the operation of classical and quantum heat machines, and providing
the later with more noteworthy advantages.

\subsection{Homogeneous energy scaling: work and heat corrections}
In the case of the homogeneous energy scaling, the working medium
is in a thermal state during the whole cycle. So we can use Wigner's
original computation to calculate the quantum corrections to the performance
of a heat machine. In 1932, Wigner calculated the quantum corrections
to the quasiprobability distribution of a M-dimensional system in
a potential $V$ at thermal equilibrium at temperature $T$ \cite{wigner1932quantum}.
After integrating the momentum, the not normalized position distribution
is given by a series expansion on $\hbar$:
\begin{equation}
\int...\int dp_{1}...dp_{M}P^{quan}(x_{1},...,x_{n};p_{1},...p_{M})=P^{clas}+\hbar^{2}P_{2}^{quan}+\hbar^{4}P_{4}^{quan}+O(\hbar^{6}),
\label{eq:wigex}
\end{equation}
where 
\begin{equation}
P^{clas}=e^{-\frac{V}{k_{B}T}}
\label{eq:pclas}
\end{equation}
is the classical phase space distribution and $P_{2n}^{quan}$ are
quantum corrections that depend on the potential and its derivatives
but not on $\hbar$. For example, the first quantum correction is
\begin{equation}
P_{2}^{quan}=e^{-\frac{V}{k_{B}T}}\left[-\frac{1}{12\left(k_{B}T\right)^{2}}\sum_{k}\frac{1}{m_{k}}\frac{\partial^{2}V}{\partial x_{k}^{2}}+\frac{1}{24(k_{B}T)^{3}}\sum_{k}\frac{1}{m_{k}}\left(\frac{\partial V}{\partial x_{k}}\right)^{2}\right].
\label{eq:p2q}
\end{equation}

At high temperatures, the Wigner function tends to the classical phase
space distribution, i.e., $\int...\int dp_{1}...dp_{M}P^{quan}(x_{1},...,x_{n};p_{1},...p_{M})\approx e^{-\frac{V}{k_{B}T}}$.
But as the temperature decreases, quantum corrections need to be included
to prevent the violation of the uncertainty principle. These corrections
change the expectation value of the energy,
\begin{equation}
\langle H\rangle_{T}=E_{clas}(V,T)+\hbar^{2}E_{2,QC}(V,T)+O(\hbar^{4}),
\label{eq:enec}
\end{equation}
where $E_{clas}(V,T)$ corresponds to the system energy predicted
by classical mechanics and $E_{2,QC}(V,T)$ is the first quantum correction
\begin{gather}
E_{2,QC}(V,T)= \notag\\
\frac{1}{\int P^{clas}}\left[\frac{1}{24k_{B}T}\int\sum\frac{1}{m_{k}}\frac{\partial^{2}V}{\partial^{2}x_{k}}P^{clas}+\int V\times P_{2}^{quan}-\frac{\int V\times P^{clas}\int P_{2}^{quan}}{\int P^{clas}}\right]
\end{gather}
where all the integrals are over all the positions, $x_{1},...,x_{M}.$
The quantum corrections to the energy affect the output of a heat
machine. For example, the heat exchanged with the hot bath, including
quantum corrections of the order $\hbar^{2}$ is
\begin{gather}
Q_{h}=Q_{h}^{clas}+\hbar^{2}\left(E_{2,QC}(V_{h},T_{h})-E_{2,QC}(V_{h},T_{c}/q)\right)+O(\hbar^{4}),
\end{gather}
where $Q_{h}^{clas}$ corresponds to the heat that a classical working
medium would exchange with the hot bath during the Otto cycle, and
the other terms are corrections that have to be included for a quantum
working medium. In the same way, we can calculate the exchanged work,
\begin{equation}
W=W^{clas}-\hbar^{2}\left(E_{2,QC}(V_{h},T_{h})-E_{2,QC}(V_{h},T_{c}/q)+E_{2,QC}(V_{c},T_{c})-E_{2,QC}(V_{c},qT_{h})\right)+O(\hbar^{4}),
\label{eq:wq}
\end{equation}
where $W^{clas}$ corresponds to the work exchange by a classical
Otto engine and the rest are the quantum corrections. As we explain later (see Eq. (\ref{eq:eneq:workq})), in order to extract work, i.e., $W\text{<0, }$ the scaling
factor of the energy levels has to comply with the following inequality:
$\frac{T_{c}}{T_{h}}<q<1$.

For 1D systems in an Otto cycle with potentials $V_{c}=a_{c}x^{2n}$
and $V_{h}=a_{h}x^{2n}$, the scaling factor between energy levels is $q=\left(\frac{a_c}{a_h}\right)^{\frac{1}{1+n}}$ \cite{zheng2014work} and the correction of order $\hbar^{2}$ to
the exchanged work can be analytically calculated. It is equal to
\begin{equation}
-\frac{\hbar^{2}\pi(2n^{2}+n-1)\csc[\frac{\pi}{2n}]}{12m(\Gamma[\frac{1}{2n}])^{2}}\left(\frac{a_{c}}{k_{B}T_{c}}\right)^{\frac{1}{n}}\left(1-\frac{1}{q}\right)\left(1-\left(\frac{T_{c}}{qT_{h}}\right)^{\frac{1}{n}}\right),
\label{eq:cor}
\end{equation}
where $\Gamma$ is the gamma function. The correction can be positive
or negative, but in the regime of work extraction, $W<0$, it is always
positive, i.e., it always reduces the work extraction.

The first quantum correction starts being relevant when the operation
temperatures are not so high and any of the Wigner functions during
the cycle deviate from the classical phase space distribution. As
temperatures decrease even more, higher order corrections should be
considered. At least for 1D systems, there are recursive formulas
that could be used to calculate any $P_{2n}^{quan}$ \cite{coffey2007wigner}, from which
the full quantum corrections to the work and heat could be calculated.
Nevertheless, this is not a practical approach and we will now introduce
a simpler strategy that is exact at any temperature.

Using the fact that the energy levels are homogeneously scaled, $E_{n}^{c}=qE_{n}^{h}$,
the expressions for the exchanged heat and work, Eqs. (\ref{eq:eneq:qh})-(\ref{eq:eneq:w}),
can be rewritten. Consider for example $Q_{h}$ (Eq. (\ref{eq:eneq:qh})).
Using the relation between the energy levels this equation becomes
\begin{gather}
Q_{h}=\sum_{n}E_{n}^{h}\left(\frac{e^{-\frac{E_{n}^{h}}{k_{B}T_{h}}}}{Z_{T_{h}}}-\frac{e^{-\frac{qE_{n}^{h}}{k_{B}T_{c}}}}{Z_{T_{c}/q}}\right)=\notag\\
\int_{T_{c}/q}^{T_{h}}\frac{d}{dT}\left(\sum_{n}E_{n}^{h}\frac{e^{-\frac{E_{n}^{h}}{k_{B}T}}}{Z_{T}}\right)dT=\int_{T_{c}/q}^{T_{h}}\frac{d\langle H_{h}\rangle_{T}}{dT}dT=\int_{T_{c}/q}^{T_{h}}C_{v}dT,\label{eq:}
\end{gather}
where we have used the fundamental theorem of calculus and $C_{v}$
is the heat capacity of a thermal state at temperature $T$ with Hamiltonian
$H_{h}$. In the same way, we can rewrite the equations for $Q_{c}$
and $W$ as: 
\begin{gather}
Q_{c}=-q\int_{T_{c}/q}^{T_{h}}C_{v}dT;\\
W=(q-1)\int_{T_{c}/q}^{T_{h}}C_{v}dT; \label{eq:eneq:workq}
\end{gather}

\begin{figure}
\begin{centering}
\includegraphics[width=1\textwidth]{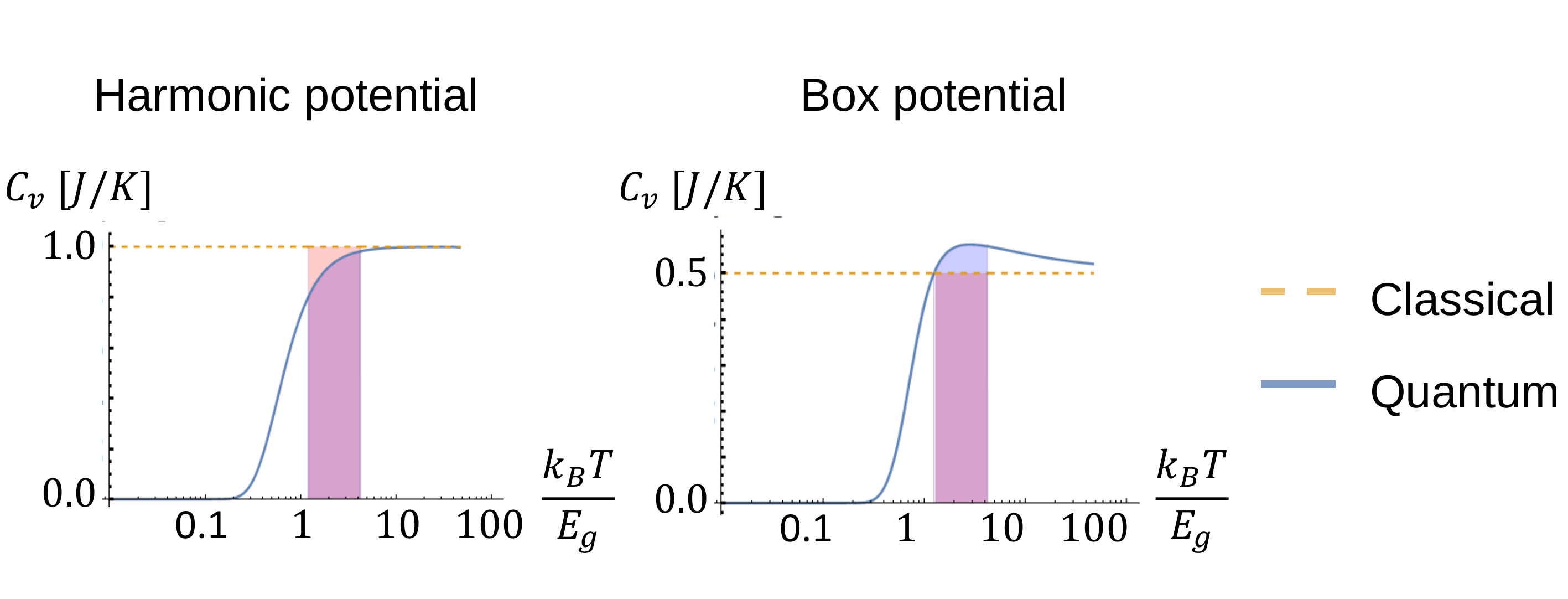} 
\par\end{centering}

\caption{Classical (yellow dashed line) and quantum (blue continuous line) heat capacities as function of the temperature normalized by the working medium ground state energy. The area below the curves, enclosed by the temperatures $T_{c}/q$ and $T_{h},$ is proportional to the work and heat exchange by a classical or  a quantum heat machine. For 1D harmonic oscillator potentials (left), the classical heat machine has a larger output than its quantum counterpart (see red area above the quantum curve). In contrast, for a particle in a 1D box (right), the output of the quantum heat machine is larger (see blue area above the classical curve).}
\label{fig:eneq:hc} 
\end{figure}

Because the heat capacity is always positive, work is extracted only if $\frac{T_{c}}{T_{h}}<q<1$. We leave the readers to prove this as an exercise.
All the information about the nature of the working medium is contained
only on the heat capacity, $C_{v}.$ If the working medium is a classical
system, classical mechanics should be used to compute $C_{v}$. If
the working medium is a quantum system, quantum mechanics should be
used to calculate $C_{v}.$ Consider for example a working medium
composed of $N$ independent harmonic oscillators at thermal equilibrium
at temperature $T$. A classical description, based on the equipartition
theorem, assigns a contribution of $\frac{k_{B}}{2}$ for each quadratic
degree of freedom of the Hamiltonian. There are two quadratic degrees
of freedom for each harmonic oscillator, \emph{x} and \emph{p}. Therefore,
classically $C_{v}^{HO,clas}=Nk_{B}$ at any $T.$ In contrast, if
we use quantum mechanics for calculating the heat capacity, we will
get $C_{v}^{HO,quan}=Nk_{B}\left(\frac{\hbar\omega}{2k_{B}T}csch\left[\frac{\hbar\omega}{2k_{B}T}\right]\right)^{2},$
which is temperature dependent (see Fig. \ref{fig:eneq:hc}-left).
Because, $C_{v}^{HO,quan}\leq C_{v}^{HO,clas},$ the heat and work
exchanged by a heat machine made of a quantum harmonic oscillator working medium is always
less than its classical counterpart, turning the classical heat machine
into a better election if we want to increase the heat or work exchange.

Nevertheless, the advantage of the classical heat machine is not a
general feature and depends on the type of working medium, which is
determined by the potentials, and the temperature range. As an example
consider a working medium composed of particles of mass $m$ in a
one dimensional box of length $L$. The quantum heat capacity is larger
than its classical counterpart at not so low temperatures, i.e., $k_{B}T\gtrsim\frac{\hbar^{2}\pi^{2}}{2mL^{2}}$
(see Fig. \ref{fig:eneq:hc}-right). Thus, for a one dimensional box
potential, a quantum heat machine may have a larger output than its
classical counterpart, i.e., extracts more work and exchanges more
heat with the thermal baths. What are the required features for a
potential in order to boost the heat machine output and how they relate
with other quantum effects, such as the Wigner function negativity,
are still open questions that should be further investigated.

Up to this point we have talked about the work and heat exchange.
But what about the thermal efficiencies, such as the engine efficiency
or the coefficient of performance of a refrigerator? Because these
quantities are rates of the exchanged work and the heat from one of
the baths,
\begin{equation}
\eta=-\frac{W}{Q_{h}}=1-q;\quad COP=\frac{Q_{c}}{W}=\frac{q}{1-q},
\label{eq:efi}
\end{equation}
they do not depend on the heat capacity, and therefore they do not
depend on the quantum or classical nature of the working medium. Thus,
$\eta^{quan}=\eta^{clas}$ and $COP^{quan}=COP^{clas}.$ The efficiencies
only depend on the potential deformation, which is characterized by
the energy levels scaling factor, $q$.
\subsection{Inhomogeneous energy scaling and the efficiency divergence}
Could the efficiency of a quantum heat machine be greater than or
at least different from its classical counterpart? Do quantum heat
machines have any fundamental advantage over their classical counterparts?
The answer to both questions is yes, but the potential transformation
should be such that the energy levels are not homogeneously scaled,
i.e., $E_{n}^{c}\neq qE_{n}^{h}.$ As we show below, not only the
performance is different, but energy quantization enables the realization
of classically inconceivable heat machines, such as those operating
with an incompressible working medium \cite{gelbwaser2017single}.

\begin{figure}
\centering{}\includegraphics[width=1\textwidth]{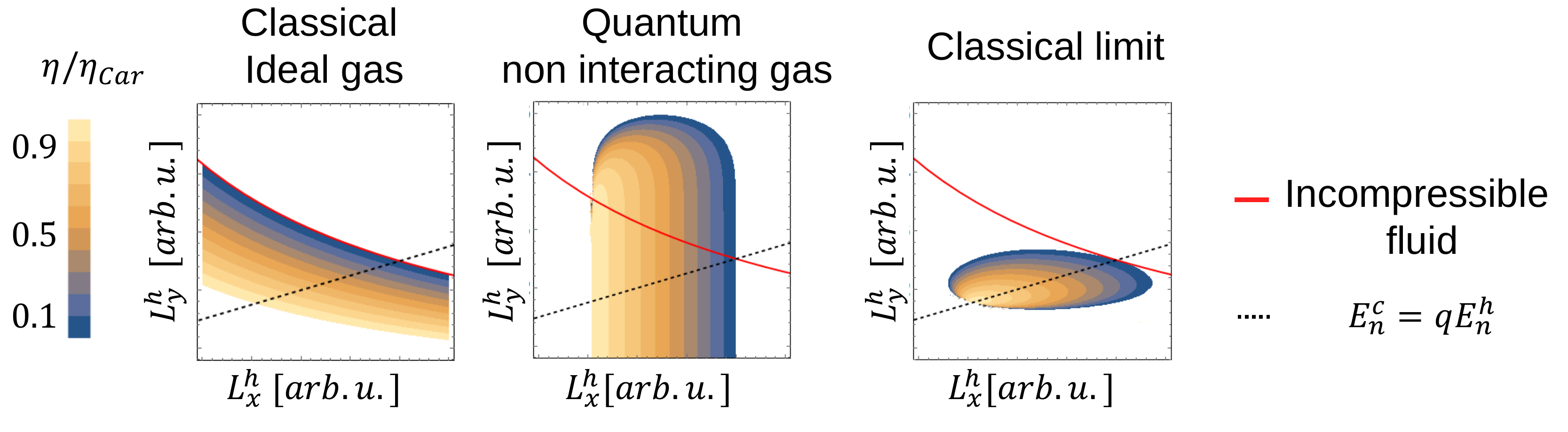}\caption{Heat engine efficiency for a 2D potential box normalized by the Carnot efficiency. At point B and C of the cycle, the box lengths are such that $L_{y}^{c}/L_{x}^{c}=0.25$ and the temperature ratio is $T_{h}/T_{c}=2$. The plots show the results for different values of the box lengths at points D and A of the cycle, i.e., $L_{x}^{h}$ and $L_{y}^{h}$. An incompressible working medium can only undergo area preserving transformations, here indicated by the red line. For these transformations there is no work extraction for a classical working medium, independently if it is an ideal gas (left) or the classical limit $\hbar\rightarrow0$ (right) of a quantum particle in a box. In contrast, a quantum working medium (center) can be highly efficient, $\eta/\eta_{Car}>0.9$, even for area preserving transformations. The dashed black line indicates transformations where the energy levels are homogeneously scaled and the quantum and classical efficiency coincide.}
\label{fig:eneq:2d} 
\end{figure}

We exemplify these effects in a particular 2D example, but a similar
analysis can be done for higher dimensions, such as 3D, or other potentials
\cite{gelbwaser2017single}. Here, we consider a working medium contained in a two dimensional
box. At points D and A of the cycle, the measurements of the box are
$L_{x}^{h}$ and $L_{y}^{h}$. At points B and C, they are $L_{x}^{c}$
and $L_{y}^{c}$ . If the working medium is incompressible, the box
transformations during the adiabatic strokes have to keep the area
constant: if the length along $x$ is increased by a factor $j$,
$L_{x}^{c}=jL_{x}^{h},$ then the size along $y$ has to be scaled
by a factor $\frac{1}{j}$, $L_{y}^{c}=\frac{1}{j}L_{y}^{h}$. These
transformations correspond to the red lines on the plots on Fig. \ref{fig:eneq:2d}.

During the adiabatic strokes, the adiabatic invariants have to be
constant. The adiabatic invariant for a classical particle of mass
$m$ and energy $E$ in a two dimensional box of area A is \cite{brown1987goodness} 
\begin{equation}
\mu=2\pi mEA,\label{eq:eneq:adin}
\end{equation}
which implies that during the adiabatic transformation	 $mEA$ is
constant. So, for a constant area adiabatic transformation, the energy
of the classical working medium does not change, precluding the work
extraction. This is true for any classical non-interacting system,
from an ideal gas to a single particle. Furthermore, the lack of work
extraction is confirmed by taking the classical limit ($\hbar\rightarrow0$)
of the quantum calculation for the work, (see Eq. (\ref{eq:eneq:w})
and Fig. \ref{fig:eneq:2d}-center).

For a quantum working medium, one has to consider the energy levels, which are given by
\begin{equation}
E_{n_{x},n_{y}}^{i}=\frac{\hbar^{2}\pi^{2}}{2m}\left(\left(\frac{n_{x}}{L_{x}^{i}}\right)^{2}+\left(\frac{n_{y}}{L_{y}^{i}}\right)^{2}\right),
\label{eq:enel}
\end{equation}
where $n_{x}$ and $n_{y}$ are integers, and $i=\{h,c\}$. For the
constant area transformations, as long as $j\neq1,$ the energy levels
scaling is inhomogeneous and the classical and quantum efficiencies
diverge. Moreover, Eq. (\ref{eq:eneq:adin}) is not an adiabatic invariant
at the quantum regime, where constant level populations is enough
to warrant adiabaticity. As stated by the quantum adiabatic theorem,
adiabaticity is achieved by performing the transformation slowly enough.
This can be realized for constant area deformations that change the
energy of the working medium. For simplicity, here we are assuming
that the quantum working medium is composed of distinguishable particles.

In contrast to the classical heat machine, a quantum heat machine
can be highly efficient, i.e., operate close to the Carnot limit,
even for constant area transformations (see red line on Fig. \ref{fig:eneq:2d}-center).
This opens the possibility of creating heat machines using incompressible
working media, which classically would be impossible. Despite the
increase in efficiency, quantum heat machines are always limited by
the Carnot bound and fully comply with the second law of thermodynamics.
The quantum model of a particle in a two dimensional box is not the
exact counterpart of a classical ideal gas heat machine. For an ideal
gas, changes on one of the directions affect the other, redistributing
the energy. In contrast, our quantum model is composed of two independent
degrees of freedom that do not interact between themselves. Its classical
limit, $\hbar\rightarrow0$, corresponds to a different performance
than the ideal gas heat machine (see Fig. \ref{fig:eneq:2d} right and left, respectively). In any case, for a classical working medium contained
in a two dimensional box, either an ideal gas or one composed of two
independent degrees of freedom, classical adiabatic invariants forbid
the work extraction for constant area transformations. This differs
from the quantum case where quantum adiabatic invariants still allow
work extraction. Further extension of this work should confirm if,
for constant area transformations, work can be extracted using a ``quantum
ideal gas'' . The gas Hamiltonian should include non-trivial interactions
among its degrees of freedom, i.e., non-linear couplings, in order
to avoid the formation of normal modes and to achieve full thermalization.
Further works should also consider the effects of  different quantum statistics, such as Fermi-Dirac or Bose-Einstein statistics.
\par
In summary, in this section we have analyzed the differences between
heat machines with a working medium governed either by classical or
quantum mechanics. The difference in their operation is determined
by the type of Hamiltonian transformation: if the energy levels are homogeneously
scaled, the work and heat exchange may diverge, but the efficiencies
are the same. In contrast, for inhomogeneous energy level scaling,
the efficiencies also diverge, allowing the extraction of work with
incompressible working fluids and opening for the realization of classically
impossible heat machines. Future research should focus on finding
other fundamental differences between the operation of classical and
quantum heat machines. In this section, we have not studied quantum
coherences or correlations. All the quantum effects that have been
considered are the basic quantum effects of a confined system: energy
quantization and the uncertainty principle. Nevertheless, as we showed,
this is enough for heat machines to have fundamentally different performances.

\section{Coherence and small action regime} 
Quantum coherence is one of the underlying principles of quantum mechanics that allows us to distinguish between classical and quantum phenomena.  
Historically, quantum coherence clarified the basic aspect of wave-particle duality in physical objects. 
Nowadays, scientists are trying to reveal the role of quantum coherence in biological, chemical, and physical systems and learn how to exploit it as a resource  \cite{plenio17,gelbwaser2017thermodynamic}.
\par  
In recent years, the role of coherence in quantum thermodynamics has been studied extensively. 
One fundamental question is how, if at all, quantum coherence can be extracted as thermodynamic work \cite{lostaglio15b,korzekwa16,skrzypczyk13,aaberg14,horodecki13,niedenzu2015performance,gelbwaser15}.  
The role of coherence in the operation of quantum thermal machines is another area of investigation. 
For example, it has been shown that noise-induced coherence can break detailed balance and enable the removel of additional power from a laser or a photocell heat engine \cite{scully11}. 
It has also been shown that coherence enhances heat flow between a quantum engine and the thermal bath, impling that coherence  plays an important role in the operation of quantum refrigerators \cite{mukamel12,mitchison15,latune18}.  
Engines subject to coherent and squeezed thermal baths have also been studied \cite{rossnagel14,manzano16,alicki2015non}. 
While such engines can exceed the Carnot efficiency or even extract work from a single bath \cite{scully03,gelbwaser2013work}, 
they do not break the second-law of thermodynamics as useful work is hidden in the bath \cite{niedenzu16}. In this context, it is important to note, that the Carnot bound is a relevant reference point only when thermal bath are considered.
\par
However, the existence of coherence in quantum thermal machines is not always beneficial for their operation. 
In \cite{seifert17} it was shown that for slow driving in the linear response of a Stirling-type engine  coherence leads to power loss.
This detraction in the performance of the engine is related to the phenomenon of quantum friction  \cite{k152,k176,zambrini14,zambrini15}. 
In this section we will discuss both positive and negative implications of coherence.
In particular, we will present a study by Uzdin et al. \cite{uzdin15,levy16b}
that reveals the thermodynamic equivalence of different types of engines in the small action regime. 
In this regime it has further been shown that coherence enhances power extraction. 
We further present a recent experiment  \cite{klatzow17} that demonstrate these two findings. 
At the end of this section we will discuss the origin of quantum friction which degrades the performance of thermal machines.  
  
\subsection{Heat machines types and the mathematical description}

\subsubsection{Heat machine types}
\label{sec:heat_mach_type}
\begin{figure}
\center{\includegraphics[width=14cm]{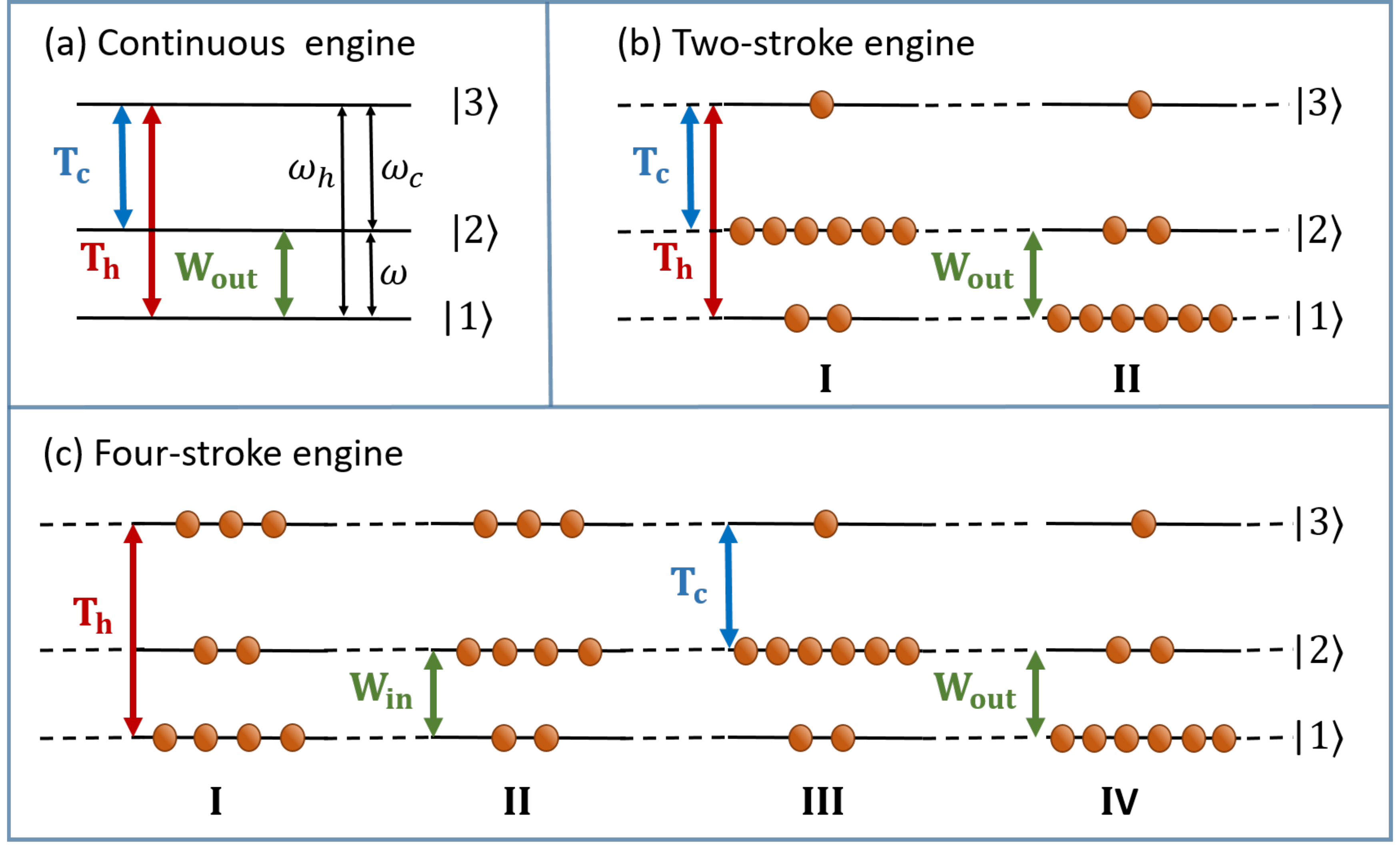}}
\caption{Scheme of three different types of heat engines: (a) a continuous engine, (b) a two-stroke engine, and (c)  a four-stroke engine.}
\label{fig1}
\end{figure}

In this section, we will consider the three most common types of heat engines: the continuous engine, the two-stroke engine, and the four-stroke engine (see Fig. \ref{fig1}).
The elementary components for assembling a quantum heat engine are two heat baths at different temperatures, $T_h$ and $T_c$, where the subscripts $h(c)$ correspond to hot(cold), such that $T_h>T_c$,
a work source, which is used for consuming/extracting energy in/out of the engine, and as the working medium, a quantum system, that couples the different components of the engine. 
\par
As was discussed in the previous chapters and in Refs. \cite{levy14a,galperin17,k85,correa14,esposito10}, the working medium can be described by different types of quantum systems. 
Here, we treat the working medium as a three-level system. 
The three-level setup was first studied by Scovil and DuBois \cite{scovil59} and is considered the pioneering work in the field. 
This model was later studied by Kosloff et al.  \cite{levy14a,k102,k169} who employed a quantum dynamical description, which reveals the significance of quantum effects in the study of thermodynamics of microscopic heat engines. 
Besides being an elementary model, the three-level engine has demonstrated  quantum signatures in nitrogen-vacancy centers experimental setup \cite{klatzow17}. 
This will be discussed in more detail in section \ref{sec:exp}.  
In Fig. \ref{fig1} the three types of quantum heat engines that are compromised of  three-level systems are described schematically. 
The Hamiltonian of the system, including the driving Hamiltonian, takes the form $H(t)=H_o + H_w(t)$ with
\beqar
\label{eq:hamiltonian}
H_o &=& \hbar \omega \ketbra{2}{2} + \hbar\omega_h \ketbra{3}{3} \\ \nonumber
H_w &=& \epsilon(t)\exp(i\omega t) \ketbra{1}{2} +\text{H.c.}.
\eeqar
Here, $H_o$ is the bare Hamiltonian, where the energy of level one is assumed to be zero.
The  system is periodically driven with the frequency $\omega=\omega_h-\omega_c$ which is in resonance with the transition frequency between the first and second levels.
The driving Hamiltonian, $H_w(t)$, is expressed after performing the rotating wave approximation (assuming $\epsilon \ll \omega$), and H.c. stands for the Hermitian conjugate.
The interaction with the bath will  be described explicitly below for both the Markovian and non-Markovian regimes.   
\par       
\textit{Continuous engines-
} are  machines in which all components of the engine are simultaneously coupled through the working medium \cite{levy12,levy14a,gelbwaser13,gelbwaser2015thermodynamics}, attaining a steady-state operation. 
The hot bath couples the first and third levels while the cold bath couples the second and third levels. 
The coupling to the heat baths generates a population inversion between the first and second levels, which is used to extract work  by amplifying a driving field connecting the two levels. 
In the weak driving limit, the condition for population inversion can be simplified to $\frac{T_h}{T_c}>\frac{\omega_h}{\omega_c}>1$. 
The device can operate as a refrigerator by simply changing the  direction of the inequality $\frac{T_h}{T_c}<\frac{\omega_h}{\omega_c}$ \cite{levy212}.
We remark that population inversion is not the only mechanism to gain power from a quantum heat engine. 
In Ref. \cite{harris16} electromagnetically induced transparency mechanism was suggested to obtain bright narrow emission light without population inversion, and this was later demonstrated experimentally \cite{zou17} with cold Rb atoms. 
\par
\textit{Two-stroke engines-}  operate in a two stage mode \cite{allahmahler08,nori07}. 
In the first stroke, the quantum system is coupled to both the cold and hot baths,
whereas in the second stroke, after the system is decoupled from the heat baths, work is extracted from the quantum system via a coupling to an external field. 
In the example of the three level system, at the end of the first stroke population inversion between the first and second levels is created, and in the second stroke this population inversion is exploited to extract useful work from the engine (see Fig. \ref{fig1}). 
\par
\textit{Four-stroke engines-} are perhaps the most familiar types of engines, as they include the Otto and Carnot engines. 
The Otto cycle, which was already introduced in Sec. \ref{sec:energy_quant} is comprised of four strokes,  two isochores and two adiabats \cite{k152,kosloff17entropy,nori14}. 
In Fig. \ref{fig1}, we describe the quantum analog of the four-stroke Otto engine.
In the first stroke, levels one and three are coupled to the hot bath, and heat flows into the working medium. 
In the second stroke, work is invested without any transfer of heat, and in the third stroke levels three and two are coupled to the cold bath, and heat flows out of the working medium.
In the last stroke, the working medium is again decoupled from the heat bath and work is extracted from the engine, completing a full cycle. 
\par     
The stroke-type engines operate repeatedly, and each cycle  takes a certain amount of time $\tau_{cyc}$. 
Defining where the cycle begins is arbitrary, as long as all four stroke are completed.
Unlike the continuous engines, the  stroke-type engines do not reach a steady state operation, but rather, they reach a limit cycle, where at the end of each cycle the state of the quantum system is the same.
The efficiencies for all of the engines described in Fig. \ref{fig1} are given by $\eta_{otto}=1-\frac{\omega_c}{\omega_h}$, which is termed the quantum otto efficiency \cite{k152,kosloff17entropy,levy14a}. 
We leave the readers to prove this as an exercise.
For simplicity, the description of the stroke-engines above is based solely on population inversion and does not require description of quantum coherence. 
In this sense, these engines describe a stochastic (classical) operation of the engines.
In the following sections we will see that the presence of quantum coherence will have an influence on the thermodynamics of these thermal devices. 
\par
The mathematical description of these engines requires tools from the theory of open quantum systems. 
While the work strokes are described using a unitary propagator $U$ which preserve the entropy of the quantum system, the coupling to the bath introduces irreversible dynamics accompanied with entropy generation. 
The dynamics of the engine is described by a completely-positive and trace-preserving map. 
In the Markovian regime this can be achieved      using the Lindblad-Gorini-Kossakowski-Sudarshan (LGKS) master equation \cite{lindblad76,gorini276}, which is described in detail in the next section.
Extending the study of quantum heat machines beyond the Markovian regime and the weak system-bath coupling limit can be preformed using  Green's functions \cite{esposito15}, the polaron transformation \cite{gelbwaser15b,segal14,cao16}, or using simulations based on the stochastic surrogate Hamiltonian \cite{k238}.
Here, we will take a different approach based on the idea of heat exchangers \cite{levy16b}, which will be described in detail in Sec. \ref{sec:non_markov}.   
\subsubsection{Markovian regime and Liouville space}
In the Markovian regime, the dynamics of the 
system (the working medium) is described by a reduced description for the density operator $\rho$ and takes the form of the LGKS Markovian master equation \cite{breuer,lindblad76,gorini276}:
\begin{equation}
i\hbar \frac{d }{dt}  \rho=L(\rho)\equiv [H_{s},\rho]+i\sum_{k}S_{k}\rho S_{k}^{\dagger}-\frac{1}{2}S_{k}^{\dagger}S_{k}\rho-\frac{1}{2}\rho S_{k}^{\dagger}S_{k}. 
\label{eq: Lind Hil}
\end{equation}
The $S_{k}$ operators depend on the system-bath coupling and on the properties of the bath, including the temperature and the correlations  \cite{breuer}. 
In following, we concentrate on thermal generators $L$ that asymptotically induce the system to evolve into a Gibbs state, $\rho_{th}=\exp(-\beta H_s)/Z$, with the inverse bath temperature $\beta=1/k_B T$ and the partition function $Z$.
This type of generator can be derived from a microscopic Hamiltonian dicription in the weak system-bath coupling limit \cite{davies74, breuer} and for a collision model in the low density limit \cite{dumcke85}.  
The necessity of a microscopic derivation was discussed in \cite{levy14b}, where it was shown that a local phenomenological description may lead to a violation of the second law of thermodynamics.
\par
To derive the results of this chapter, we analyze the dynamics in the extended Liouville space. 
In this space, any matrix representation of an operator acting in a Hilbert space is mapped to a vector. 
That is, for a general matrix $B$ acting in Hilbert space:   $B_{N\times N} \rightarrow\ket{B}_{1\times N^2}$. 
Given this index mapping, the master equation (\ref{eq: Lind Hil}) in Liouville space reads:
\begin{equation}
i\hbar \frac{d }{dt}\ket{\rho} \equiv \mc L \ket{\rho} \equiv (\mc H_{s}+\mc L^d)\ket{\rho}.
\label{eq: Lind Lio}
\end{equation}
The super-operator $\mc H_{s}$ is represented by a Hermitian $N^{2}\times N^{2}$ matrix that corresponds to the first term on the right hand side of (\ref{eq: Lind Hil}), and the super-operator $\mc L^d$ is a non-Hermitian $N^{2}\times N^{2}$ matrix that corresponds to the dissipative terms in (\ref{eq: Lind Hil}).  
In this chapter, we use calligraphic letters to describe super-operators acting in Liouville space and ordinary letters for operators acting on a Hilbert space.
We also chose a specific map known as the ``vec-ing''. 
More details on the Liouville-space representation of quantum mechanics are presented in Box 1.
\par
The LGKS operators in Hilbert space that describes the coupling of the system to the hot bath are expressed as
\beqar
S_1^h &=& \sqrt{\gamma_h}e^{-\hbar\omega_h\beta_h}\ketbra{3}{1},\\ \nonumber
S_2^h &=& \sqrt{\gamma_h}\ketbra{1}{3},
\eeqar
and those to the cold bath as
\beqar
S_1^c &=& \sqrt{\gamma_c}e^{-\hbar\omega_c\beta_c}\ketbra{3}{2},\\ \nonumber
S_2^c &=& \sqrt{\gamma_c}\ketbra{2}{3}.
\eeqar
These operators form the dissipators $\mathcal{L}_h$ and $\mathcal{L}_c$ respectively. 
In the weak coupling limit, the parameters $\gamma_h$ and $\gamma_c$ are given by the Fourier-transforms of the baths correlation functions \cite{breuer}. 
Without any external driving, the coupling to the baths  will lead  asymptotically to a Boltzmann factor ratio of the populations, $p_3/p_1=e^{-\hbar\omega_h\beta_h}$ and $p_3/p_2=e^{-\hbar\omega_c\beta_c}$. 
However, when driving is incorporated to the process,  the behavior of the state is changing, and different engine-types will differ from one another.
With these building blocks a full description in the Markovian regime of the different engines can be obtained. 
\cleardoublepage
\begin{framed}
\begin{center}
\textbf{Box 1 : Liouville-space representation of quantum mechanics}
\end{center}
Quantum dynamics is traditionally described in Hilbert space. However, it is convenient,  particularly, for open quantum systems, to introduce an extended space where the density operator is represented by a vector and the time evolution of a quantum system is generated by a Schr$\ddot{\text{o}}$dinger-like equation. 
This space is usually referred to as Liouville space \cite{mukamel1999principles}. 
We denote the ``density vector'' by $\ket{\rho}\in\mathbb{C}^{1\times N^{2}}$,
which is obtained by reshaping the density matrix $\rho$ into a larger
vector with index $\alpha\in\{1,2,....N^{2}\}.$ The one-to-one mapping of the two matrix indices into a single vector index $\{i,j\}\to\alpha$ is arbitrary, but has to be used consistently. 
In general, the vector $\ket{\rho}$ is not normalized to unity. Its norm is equal to the purity, $\mc P=\text{tr}(\rho^{2})=\braket{\rho}{\rho}$, where $\bra{\rho}=\ket{\rho}^{\dagger}$. 
The equation of motion of the density vector in Liouville
space follows from
\beq
 d_{t}\rho_{\alpha}=\sum_{\beta}\rho_{\beta}\partial(d_{t}\rho_{\alpha})/\partial\rho_{\beta}.
\eeq
Using this equation, one can verify that the dynamics of the density
vector $\ket \rho$ is governed by a Schr$\ddot{\text{o}}$dinger-like equation in the
new space, 
\begin{equation}
i\hbar \partial_{t}\ket{\rho}=\mc L\ket{\rho},
\label{eq: schrodinger eq}
\end{equation}
where the super-operator $\mc L \in\mathbb{C}^{N^{2}\times N^{2}}$
is given by 
\begin{equation}
\mc L_{\alpha\beta}=i\hbar\frac{\partial(d_{t}\rho_{\alpha})}{\partial\rho_{\beta}}.
\label{eq: Hr form}
\end{equation}
\par
A particularly useful index mapping is the ``vec-ing'' maping  \cite{machnes14,roger1994topics,am15} that provides a simple form for $\mc L$ in terms of original Hilbert-space Hamiltonian and Lindblad
operators. 
In this mapping, the density vector $\ket \rho$ is ordered row by row, i.e., $\alpha$ = col+$N$(row-1), and the following relations hold:
\beqar
[H,\rho] &\rightarrow& \left(  H\otimes I - I \otimes H^{T} \right)\ket{\rho}\\ 
\nonumber
A\rho A^{\dagger} &\rightarrow&  \left( A\otimes A^* \right)\ket{\rho}\\ 
\nonumber
A^{\dagger} A \rho  &\rightarrow&  \left( A^{\dagger}A  \otimes I\right)\ket{\rho}\\ \nonumber
 \rho A A^{\dagger} &\rightarrow&  \left(  I \otimes \left(A^{\dagger}A\right)^T  \right)\ket{\rho},
\eeqar
where $^T$ denotes the transpose and $^*$ denotes the complex conjugate.
Using these relations the generator $\mc L$  can easily be expressed in a matrix form. 
The dynamical map generated by the Lindblad super-operator can also be expressed in a matrix form $\Lambda =\exp({-i\mc L t/ \hbar})$. 
This matrix has a single eigenvalue which is equal to one and it's eigenvector is associated with the  stationary state of the system. 
\par
For standard thermalization dynamics, this state is the Gibbs (thermal) state with the temperature of the bath. 
The inner product of two operators  $A$ and $B$ in Hilbert space is mapped to Liouville-space as
\beq
\label{eq:inner_prod}
tr(A^{\dagger}B)\rightarrow \braket{A}{B}.
\eeq  
Since $\rho$ is Hermitian, (\ref{eq:inner_prod}) implies that the expectation value of an operator $A$ in Liouville space is given by the inner product of $\rho$ and $A$,
\beq
\ave A = \braket{\rho}{A}.
\eeq 
The dynamics of the expectation value $\ave A$ can then be expressed as
\beq
\frac{d}{dt}\ave A = \frac{i}{\hbar}\bra{\rho}\mc L^{\dagger}\ket{A}+\braket{\rho}{\partial_t A}.
\eeq
Another useful relation for the Hamiltonian part 
\beq
\mathcal{H}\ket{H}=\bra{H}\mc H=0,
\eeq
where we mapped $H\rightarrow \ket{H}$ and $[H,\cdot]\rightarrow\mathcal{H}$. 
This relation follows from the fact that the Hamiltonian commutes with itself. 

\end{framed}
\subsubsection{Non-Markovian and strong coupling regime}
\label{sec:non_markov}
Generally, treating non-Markovian dynamics and a strong system-bath coupling can be challenging. 
Here we adopt the idea developed in \cite{levy16b} of heat exchangers that captures the effects of non-Markovianity and strong coupling on the quantum thermal-machines equivalence principle which will be discussed in Sec. \ref{sec:qtme}.   
\par  
Heat exchangers are widely used in engineering to pump heat out of a system. For example, computer chips interact strongly with a metal that conducts the heat and then is cooled by the surrounding air.
In the context  of the engines described in Fig. \ref{fig1}, the quantum system interacts with  heat exchangers that are modeled  by two level particles.
There are $N_h$ and  $N_c$ particles in the hot and the cold heat exchangers, respectively. 
We assume that in each stroke, the working medium interacts with a single particle of the heat exchangers.
Before the interaction, each of the particles in the exchangers is in a thermal state with temperature $T_{h,c}$. 
During the interaction, heat is exchanged with the quantum system and the states of the particle and the working medium change.
After the interaction, the particle relaxes back to its original thermal state  via a coupling to a thermal bath. 
Thus, each particle will interact with the working medium cyclically with a period $N_{h,c}\tau_{cyc}$. 
\par
The advantage of the above description is that it is independent of the properties of the thermal bath (the interaction can be described in the weak coupling and Markovian limits).
On the other hand, the particles may interact strongly with the working medium and in a non-Markovian manner, imposing the effects of strong and non-Markovianity on the equivalence principle.  
\par
In this scheme, the work source is modeled by a set of qubits or qutrits interacting with the working medium. 
After a complete cycle of the engine, energy is stored in the work repository which can later be extracted, similar to the function of a battery. 
The concepts of quantum batteries and quantum flywheels, where energy is stored and extracted from internal degrees of a quantum system  were studied in different contexts \cite{levy16,levy16b,alicki13,binder15,marti15,korzekwa15}
\par
When the work repository is quantized, the exchange of energy with the working medium cannot necessarily be considered as pure work, since entropy may be generated in the battery and the working medium \cite{levy16, levy16b,woods15,gelbwaser2014heat,gelbwaser2013workqp}.
This phenomenon is ignored when the work repository is modeled as a semi-classical field.
The field is typically considered large enough that the entropy generated in the field is negligible, and since its operation on the quantum system is unitary the entropy of the  quantum system does not change as well. 
This generation of entropy in the battery can be resolved using a feedback scheme \cite{levy16} or by applying a specific procedure that guarantees that entropy will be produced in the working medium while the entropy of the battery is not changing or even reduced (super-charging) \cite{levy16b}. 
\par
The interaction of the working medium with the work repository and the heat exchangers can be modeled by unitary dynamics evolving from a Hamiltonian description. 
The initial states of the heat exchangers, the work repository, and the working medium in each cycle are uncorrelated, i.e., $\rho_{tot}(0)=\rho_c \otimes \rho_h\otimes \rho_w\otimes \rho$. Here $\rho$ is the engine (working medium) state and $\rho_{c,h,w}$ are the heat exchangers and work repository states. 
The coupling between the engine and the rest of the particles takes the form
\beq
H_{int}=\sum_{k=c,h,w}f_k(t)H_{ok},
\eeq
where $f_k(t)$ is a periodic function that control the timing of the interaction.
We further require that the interactions will be energy conserving. 
Thus the energy in the exchangers, the work repository, and the engine are not effected by $H_{ok}$ but only redistributed. 
Mathematically this requirement is translated to
$[H_{ok},H_o+H_k]=0$, where the index $k$ stands for $c,h,w$, and $H_o$ is defined by Eq. (\ref{eq:hamiltonian}).
The interaction Hamiltonian takes the form, 
\beq
H_{ok}=a_k^{\dagger}a_{ok}+ a_ka_{ok}^{\dagger},
\eeq
where $a_k$ is the annihilation operator of the $k$ particle of the exchangers and battery, and $a_{ok}$ is the annihilation operator of the $k$ manifold of the working medium (the three-level system, see Fig. \ref{fig1}). 
This type of interaction generates a partial or full swap between the $k$ manifold of the engine and the corresponding $k$ particle of the exchangers and battery. 
Since the swap operation may significantly change  both the states of the heat exchangers particles and the engine and form quantum correlations, this simplified model captures non-Markovian and strong coupling effects.  
\subsection{Quantum thermal machines equivalence}
\label{sec:qtme}
A peculiar phenomenon in the thermodynamic behavior of the different engine-types described above  occurs in a quantum regime where the action over a cycle is small compared to $\hbar$ \cite{uzdin15}.
In this regime of operation, the four-stroke, two-stroke, and continuous  engines have the same thermodynamic properties for both transient and steady state operation.
At the end of each cycle, the power, heat and efficiency becomes equivalent. 
This phenomenon can be traced back to the coherent mechanisms of the engines that become dominant in the small action regime.
The action is defined as the integral over a cycle time of the spectral norm of the generator of the dynamics:
\beq
\label{eq:actio_norm}
s=\int_0^\tau \parallel \Lc(t) \parallel dt.
\eeq 
The spectral norm (operator norm) is simply given by $\parallel \Lc \parallel = \text{max}\sqrt{\text{eig}(\Lc^{\dagger} \Lc)}$. 
For a non-Hermitian operator, this magnitude is the largest singular value of the operator $\Lc$.
The action norm Eq. (\ref{eq:actio_norm}) has been used before to obtain quantum speed limits and distance bounds for quantum states \cite{uzdin16,lidar08} and the distance between  protocols in the context of quantum control \cite{levy17}. 
As an example \cite{uzdin15}, the action norm limits the maximal state change during time $\tau$, such that $\text{max}\left(\parallel \ket{\rho(\tau)}-\ket{\rho(0)}\parallel \right) \leqslant s/\hbar$.
\par
The derivation of the equivalence is based on the Strang decomposition \cite{jahnke00,strang68} for two non-commuting operators $\mathcal{A}$ and $\mathcal{B}$,
\beq
\label{eq:strang}
e^{(\mathcal{A}+\mathcal{B})dt}= e^{\half\mathcal{A}dt}e^{\mathcal{B}dt}e^{\half\mathcal{A}dt}+O\left(\bar{s}^3 \right)\cong e^{\half\mathcal{A}dt}e^{\mathcal{B}dt}e^{\half\mathcal{A}dt}.
\eeq 
Here we define $\bs \equiv s/\hbar$, which is the norm action $s=(\parallel \mathcal{A}\parallel + \parallel \mathcal{B} \parallel)dt$ divided by $\hbar$  that must  be small parameter for the expansion to hold, i.e. $\bs \ll 1$ \cite{uzdin15}.

\subsubsection{Equivalence in the Markovian regime}
\label{sec:equiv_markov}
To derive the engines equivalence we start with the dynamical description of the continuous engine. 
The following calculations are carried out in a rotating frame, according to the transformation $\mathcal{U}=e^{-i\mathcal{H}_o t/\hbar}$. 
Since $\Lch$ and $\Lcc$ commute with $\mathcal{H}_o$ the transformation will not affect these generators. 
The interaction Hamiltonian is now time independent $H_w\rightarrow \tilde{H}_w = \epsilon\ketbra{1}{2}+\text{H.c.} $ (see Eq. (\ref{eq:hamiltonian})), and the generator of the dynamics reads,
\beq 
\tilde{\mathcal{L}}=\Lch + \Lcc +\tilde{\mathcal{H}}_w.
\eeq
We chose the cycle time $\tau_{cyc}=6m\tau_d$, where $\tau_d$ is the external drive cycle and $m\in\mathbb{Z}^+$.
The propagator of the continuous engine over time $\tau_{cyc}$ is
\beq
\label{eq:cont}
\Lambda^{cont}=
\exp \left[-i (\Lch+\Lcc+\tilde{\mathcal{H}}_w)\tfrac{\tau_{cyc}}{\hbar} \right].
\eeq 
Applying the Strang splitting Eq. (\ref{eq:strang}) on $\Lch+\Lcc$ we obtain the two-stroke propagator over one cycle,
\beq
\label{eq:2st}
\Lambda^{2st}= \exp \left[-i\tfrac{3}{2}(\Lch+\Lcc)\tfrac{\tau_{cyc}}{3\hbar} \right]
\exp \left[-i3\tilde{\mathcal{H}}_w\tfrac{\tau_{cyc}}{3\hbar}\right]
\exp \left[-i\tfrac{3}{2}(\Lch+\Lcc)\tfrac{\tau_{cyc}}{3\hbar} \right].
\eeq
Here we rescaled the cycle time and the couplings to the bath and the external field such that the total cycle time will remain $\tau_{cyc}$. Moreover, we set the fraction time of the work stroke to be $1/3$ of the cycle time, in agreement with the experiment \cite{klatzow17} described in Sec. \ref{sec:exp}.
We can repeat this procedure and obtain the four-stroke propagator over one cycle. 
First we split $\Lcc$ in Eq. (\ref{eq:cont}) and then split $\mathcal{H}_w$,
\beq
\label{eq:4st}
\Lambda^{4st}= \exp \left[-i3\Lcc\tfrac{\tau_{cyc}}{6\hbar} \right]
\exp \left[-i3\tilde{\mathcal{H}}_w\tfrac{\tau_{cyc}}{6\hbar}\right]
\exp \left[-i3\Lch\tfrac{\tau_{cyc}}{3\hbar} \right]
\exp \left[-i3\tilde{\mathcal{H}}_w\tfrac{\tau_{cyc}}{6\hbar}\right]
\exp \left[-i3\Lcc\tfrac{\tau_{cyc}}{6\hbar} \right].
\eeq   
Also here, the cycle time and couplings are rescaled to maintain a similar $\tau_{cyc}$ for all engines.
According to Eq. (\ref{eq:strang}), given $\bs \ll 1$, all  engine-types propagators over a cycle are equivalent to order $O(\bs^3)$, that is
\beq
\label{eq:prop_equiv}
\Lambda^{cont} \cong \Lambda^{2st} \cong \Lambda^{4st}.
\eeq 
The equivalence holds also for the average  work and heat over a cycle:
\beqar
\label{eq:WQ_equiv}
W^{cont} &\cong & W^{2st} \cong  W^{4st}\\ \nonumber
Q_{c,h}^{cont} &\cong & Q_{c,h}^{2st} \cong  Q_{c,h}^{4st}.
\eeqar 
As heat and work are process-dependent, and since  the states of the different engines differ significantly from one another during the cycle, Eq. (\ref{eq:WQ_equiv}) should be proved.
The rigorous proof can be found in \cite{uzdin15} and is based on the symmetric rearrangement theorem. 
Here we will explicitly show how work equivalence  can be derived for the continuous and the two-stroke engines.
In steady state, the work performed over one cycle of the continuous engine is given by the steady state power multiplied by the cycle time \cite{uzdin15},
\beq
\label{eq:W_cont}
W^{cont}=\frac{-i}{\hbar}\bra{H_o}\tilde{\mathcal{H}}_w\ket{\tilde{\rho}_s}\tau_{cyc}.
\eeq
Here $\tilde{\rho_s}$ is the steady state density matrix of the continuous engine in the rotating frame.
The work output over a single cycle of the two-stroke engine is simply given by the  energy difference between the end and beginning of the work stroke.
Assuming that the engine operates in the limit cycle and that the cycle (\ref{eq:2st}) starts at a state $\ket{\tilde{\rho}(t_o-\tau_{cyc}/2)}$, we have, $\ket{\tilde{\rho}(t_o-\tau_{cyc}/2)}=\ket{\tilde{\rho}(t_o+\tau_{cyc}/2)}$.
The work in this cycle is then given by
\beq
\label{eq:W_2st}
W^{2st}=\braket{H_o}{\tilde{\rho}\left(t_o+\tfrac{\tau_{cyc}}{6}\right)}-\braket{H_o}{\tilde{\rho}\left(t_o-\tfrac{\tau_{cyc}}{6}\right)}=\tfrac{-i}{\hbar}\bra{H_o}\tilde{\mathcal{H}}_w\ket{\tilde{\rho}(t_o)}\tau_{cyc}+O(\bs^3).
\eeq 
The second equality in Eq. (\ref{eq:W_2st}) follows from the fact that in the small action regime we can expand the states $\ket{\tilde{\rho}(t_o\pm \tau_{cyc}/6)}=\ket{\tilde{\rho}(t_o)}\mp i3\tilde{H}_w \tfrac{\tau_{cyc}}{6\hbar}\ket{\tilde{\rho}(t_o)}+O(\bs^2)$.  
Because of symmetry, the second order term vanishes and we are left with correction of the order $O(\bs^3)$. 
Using Eq. (\ref{eq:prop_equiv}), we have $\ket{\tilde{\rho}\left(t_o\pm \tau_{cyc}/2\right)}=\ket{\tilde{\rho}_s}+O(\bs^3)$, and  $\ket{\tilde{\rho}(t_o\pm \tau_{cyc}/2)}=\ket{\tilde{\rho}(t_o)}\mp i\tilde{\mathcal{L}} \tfrac{\tau_{cyc}}{2\hbar}\ket{\tilde{\rho}(t_o)}+O(\bs^2)$,  which implies $\ket{\tilde{\rho}(t_o)}=\ket{\tilde{\rho}_s}+O(\bs^2)$. Inserting this last relation into the right hand side of Eq. (\ref{eq:W_cont}), we obtain $W^{cont}=W^{2st}+O(\bs^3)$.
In a similar manner, one can show the equivalence of the heat which is defined as
\beq
\label{eq:Q_cont}
Q_{c,h}^{cont}=\frac{-i}{\hbar}\bra{H_o}\mathcal{L}_{c,h}\ket{\tilde{\rho}_s}\tau_{cyc}.
\eeq
The proof above assumes operation in the limit cycle. However, note that the equivalence holds also for transients and can be shown to hold for the four stroke engine without explicit calculation. 
To show this, the symmetric rearrangment theorem \cite{uzdin15} should be applied. 
\par
The average power and heat flow are given by $ P=W/\tau_{cyc}$ and $J_{c,h}=Q_{c,h}/\tau_{cyc}$. 
The power and heat flows of the continuous engine is independent of the cycle time $\tau_{cyc}$. Thus, according to the proof of the equivalence for the work and heat, the power and heat flow of the two-stroke and four-stroke engines will deviate quadratically from the power of the continuous one,
\beq
\label{eq:power_equiv}
P^{i}=P^{cont}+\alpha_i\bs^2 \quad \text{for} \quad i=2st, 4st.
\eeq 
This result can be observed in Fig. \ref{fig:exp}b and will be discussed in detail in Sec. \ref{sec:exp}.
\subsubsection{Equivalence in the strong-coupling and non-Markovian regime}
The thermodynamic equivalence of heat machines can be extended  to the non-Markovian and strong coupling regime.
Adapting the setup introduced in Sec. \ref{sec:non_markov} and applying  the Strang decomposition Eq. (\ref{eq:strang}), one can show the equivalence principle for small action in a similar manner to the Markovian regime. 
Yet, the equivalence work and heat in this setup is of the order of $O(\bs^4)$, instead of $O(\bs^3)$ as it is in the Markovian regime. 
The dynamics of the engine and its interaction with the heat exchangers and the battery are described by a unitary transformation. 
By choosing the energy gaps in the engine manifold $k$ to match  those of the $k$-th heat exchanger and battery,  we have the condition $[H_o+\sum_{k=h,c,w}H_k,H_{int}]=0$. 
This condition implies that transformation to a rotated frame according to $U_0=\exp[-i(H_o+\sum_{k=h,c,w}H_k)t/\hbar]$ will not affect $H_{int}$. Thus, in the rotated frame, the propagators in Hilbert space of  continuous\footnote{The proper name for this type of engine would be the simultaneous engine, as all the interactions occur within one stroke. After this stroke, some energy is stored in the battery and the engine will subsequently interact with a new set of particles. However, for consistency with previous sections we will refer to this engine as continuous.},
 two-stroke, and four-stroke engines over a cycle time read:
\beqar
U^{cont} &=& \exp[-i(H_{oh}+H_{oc}+H_{ow})\tfrac{\tau_{cyc}}{\hbar}]\\ \nonumber
U^{2st} &=& \exp[-i\tfrac{3}{2}H_{ow}\tfrac{\tau_{cyc}}{3\hbar}]
\exp[-i3(H_{oh}+H_{oc})\tfrac{\tau_{cyc}}{3\hbar}]
\exp[-i\tfrac{3}{2}H_{ow}\tfrac{\tau_{cyc}}{3\hbar}]\\ \nonumber
U^{4st} &=& \exp \left[-i3H_{oc}\tfrac{\tau_{cyc}}{6\hbar} \right]
\exp \left[-i3H_{ow}\tfrac{\tau_{cyc}}{6\hbar}\right]
\exp \left[-i3H_{oh}\tfrac{\tau_{cyc}}{3\hbar} \right]
\exp \left[-i3H_{ow}\tfrac{\tau_{cyc}}{6\hbar}\right]
\exp \left[-i3H_{oc}\tfrac{\tau_{cyc}}{6\hbar} \right].
\eeqar
Applying the Strang decomposition in a similar manner to Sec. \ref{sec:equiv_markov} results in
\beq
\label{eq:U_equiv}
U^{cont}(\tau_{cyc})\cong U^{2st}(\tau_{cyc}) \cong U^{4st}(\tau_{cyc})
\eeq   
up to order $O(\bs^3)$.
If the engines start in the same initial condition, then their states at time $m\tau_{cyc}$ for $m\in\mathbb{Z}^+$ will differ at most by $O(\bs^3)$, while at other times, they will differ in the strongest order possible  $O(\bs)$. 
The heat and the work over one cycle are given by the energy of the heat  exchanger particle and the energy stored in the battery after this cycle,
\beqar
\label{eq:Q_and_W}
W^j&=&tr\left[\left(U^j(\tau_{cyc}) \rho_{tot}(0)U^j(\tau_{cyc})^{\dagger} -\rho_{tot}(0)\right)H_w \right] \\ \nonumber
Q^j_{c,h}&=&tr\left[\left(U^j(\tau_{cyc}) \rho_{tot}(0)U^j(\tau_{cyc})^{\dagger} -\rho_{tot}(0)\right)H_{c,h} \right] \quad j=cont, 2st, 4st. \\ \nonumber
\eeqar
Equations (\ref{eq:U_equiv}) and (\ref{eq:Q_and_W})  imply immediately, without the need of the symmetric  rearrangement theorem, the  equivalence of heat and work for the different heat engine types.
\par
Since at every cycle the engine interacts with  new particles  of the battery and the heat exchangers, and no initial correlation is present, the equivalence of heat and work hold up to order $O(\bs^4)$.
The $O(\bs^3)$ corrections contribute only to inter-particle coherence  generation and not to population changes.
The inter-particle coherence should be distinguished from the single particle coherence, which is manifested by local operations on the single particles. 
The inter-particle coherence represents the interaction between degenerate states. In the model describe above, we have three pairs of degenerate states $\{\ket{0_w 2}, \ket{1_w 1}  \}$,  $\{\ket{0_h 3}, \ket{1_h 1}  \}$, and $\{\ket{0_c 3}, \ket{1_c 2}  \}$, which are essential for the engine operation. 
Suppression of these coherences will lead to a Zeno effect, where the engine will not evolve in time.
\par    
Here, we assumed that the initial states of all the engines are the same, which implies that the equivalence also holds in the transient dynamics.  However, the equivalence holds also in the limit cycle regardless of the initial state of the engines. 
To show this one should look at the reduced state of one of the engine types at the limit cycle and apply the evolution operator of a different engine and see how the state changes.

\subsection{Quantum-thermodynamic signatures}
Much like in  quantum information theory, where  entanglement witness are identified in order to distinguish between entangled and separable states, it is desirable to identify quantum thermodynamic signatures in the operation of quantum thermal machines.
We define a quantum-thermodynamic signature as a thermodynamic measurement, such as power or heat flow, that confirms the presence of  quantum effects, such as coherence or quantum correlations, in the operation of the device.   
In this section, we will focus on the presence of coherence (interference) in the operation of  quantum heat engines  and its signature \cite{uzdin15}.
\par
A quantum-thermodynamic signature is constructed by setting a bound on a thermodynamic measurement of a classical engine.
In this section we define a classical engine as a device that can be fully described by its population dynamics.   
Considering, for example, the two-stroke engine described in \ref{sec:equiv_markov}, the work done in the work stroke is given by
$W=\bra{H_o}\exp[-\tfrac{i}{\hbar}\tilde{\mathcal{H}}_w\tau_w]\ket{\tilde{\rho}}-\braket{H_o}{\tilde{\rho}} $, where $\tau_w$ is the work stroke duration.
Splitting the state into the diagonal (population) contribution and to off-diagonal (coherence) contribution such that $\ket{\tilde{\rho}}=\ket{\tilde{\rho}_{pop}}+\ket{\tilde{\rho}_{coh}}$, we can express the work as
\beq
\label{eq:split_work}
W =  
\bra{H_o} \sum_{n=1} \frac{(-\tfrac{i}{\hbar}\tilde{\mathcal{H}}_w\tau_w)^{2n}}{2n!} \ket{\tilde{\rho}_{pop}} + 
\bra{H_o} \sum_{n=1} \frac{(-\tfrac{i}{\hbar}\tilde{\mathcal{H}}_w\tau_w)^{2n-1}}{(2n-1)!} \ket{\tilde{\rho}_{coh}}.
\eeq
The above splitting is possible because of two reasons. 
First, note that  $\tilde{H}_w$ in Hilbert space  only has off-diagonal terms in the energy basis of $H_o$. 
This implies that in Liouville space, the super-operator and the state will have the structure \footnote{In the more general case where $\tilde{H}_w$  also has diagonal terms, there will be an additional entry to the matrix that does not vanish, (i.e.,  $\tilde{\mathcal{H}}_w[2,2]\neq 0$) and couples coherences to coherences. }
\beq
\label{eq:Hw_struc}
\tilde{\mathcal{H}}
_w = \begin{pmatrix}
0 & h \\
h^{\dagger} & 0
\end{pmatrix} \quad \quad
\ket{\tilde{\rho}} = \begin{pmatrix}
pop \\ coh
\end{pmatrix}.
\eeq
This means that $\tilde{\mathcal{H}}_w$ couples only coherences to populations and populations to coherences. 
Second, since in Hilbert space $H_o$ is diagonal,   $\bra{H_o}$ projects the state on the population space in Liouville space. 
Thus, contributions to the work will come from odd powers of $\tilde{\mathcal{H}}_w$ operating on coherences and from even powers operating on populations.    
\par
To construct classical (stochastic) engines based on our previous description of the two-stroke and the four-stroke quantum engines, we introduce a dephasing operator $\mathcal{D}$ that will eliminate  coherences in the engine.
This is achieved by operating with $\mathcal{D}$ at the beginning and at the end of each stroke. 
We require that the operator $\mathcal{D}$   dephase the system in the  energy basis such that it does not change the energy population. This is referred as pure dephasing, which means that before and after the operation of $\mathcal{D}$, the energy of the engine and the population in the energy basis are the same.
Conceptually this operator can be described  as a projection operator on the population      space $\mathcal{D}=\ketbra{pop}{pop}$.
In practice, this operator can be expressed in Hilbert space, for example, as $D(\rho)=-\eta[H_o,[H_o,\rho]]$ with  the dephasing rate $\eta>0$. 
This term  may arise in different physical scenarios  \cite{milburn91,levy17,diosi88}.
Since $\mathcal{D}$ commutes with $\mathcal{L}_{h,c}$,  the effect of $\mathcal{D}$ on the dynamics and the work is reduced to its influence on the unitary strokes. 
The work of the stochastic engine can now be expressed as $ W^{stoch}=\bra{H_o}\mathcal{D}(\exp[-\tfrac{i}{\hbar}\tilde{\mathcal{H}}_w\tau_w]-\mathbb{I})\mathcal{D}\ket{\tilde{\rho}} = \bra{H_o}(\exp[-\tfrac{i}{\hbar}\tilde{\mathcal{H}}_w\tau_w]-\mathbb{I})\ket{\tilde{\rho}_{pop}} $, where we used the relation $\bra{H_o}\mathcal{D}=\bra{H_o}$.
In the small action regime the leading order to the work is quadratic in $\mathcal{H}_w$, 
\beq
\label{eq:w_stoch}
W^{stoch}= -\frac{\tau_w^2}{2\hbar^2}\bra{H_o}\tilde{\mathcal{H}}_w^2\ket{\tilde{\rho}_{pop}}+O(\bs^4).
\eeq 
Energy leaving the system is always negative, hence we will bound the absolute value of the work produced by the engine.
Applying H\"{o}lder's inequality  we obtain  
$|\bra{H_o}\tilde{\mathcal{H}}_w^2\ket{\tilde{\rho}_{pop}}| \leq \parallel\bra{H_o}\tilde{\mathcal{H}}_w^2\parallel_{\infty}\parallel\tilde{\rho}_{pop}\parallel_1$.
Since $\parallel \tilde{\rho}_{pop}\parallel_1=1$, the upper bound on the work and power is independent of the state of the engine and is determined solely by the configuration of the engine,
\beqar
\label{eq:up_bound}
|W^{stoch}| &\leq & \tfrac{\tau_w^2}{2\hbar^2} \parallel\bra{H_o}\tilde{\mathcal{H}}_w^2\parallel_{\infty} \\ \nonumber
|P^{stoch}| &\leq & \tfrac{d^2\tau_{cyc}}{2\hbar^2} \parallel\bra{H_o}\tilde{\mathcal{H}}_w^2\parallel_{\infty}.
\eeqar  
Here $d=\tfrac{\tau_w}{\tau_{cyc}}$ is the partial duration of the work stroke.
The upper bound (\ref{eq:up_bound}) is tighter than the one introduced in \cite{uzdin15} as the norm $\parallel\cdot \parallel_{\infty}\leq \parallel\cdot\parallel_{2}$.
For the Hamiltonian (\ref{eq:hamiltonian}) the two-stroke stochastic engine upper bound is then given by 
\beq
\label{eq:up_bound2}
|P^{stoch}| \leq \hbar \omega |\epsilon |^2 d^2 \tau_{cyc}.
\eeq
\par
Any power measurement that exceeds the stochastic bound $|P|>|P^{stoch}|$ is considered  a quantum-thermodynamic signature, indicating the existence of interference in the engine (see Fig. \ref{fig:exp}a). Thus, exceeding the classical bound is a sufficient (but not necessary) condition that the engine incorporates coherence in its operation. 
Two immediate results can be derived from the above construction. First, continuous engines with the Hamiltonian structure (\ref{eq:Hw_struc}), only  have a coherent work extraction mechanism, which is a result of Eq. (\ref{eq:W_cont}) and the splitting (\ref{eq:split_work}). 
Thus, complete dephasing  nulls the power.
Second, given the same thermodynamic resources, in the small action regime, a coherent quantum engine will outperform a stochastic (classical) engine, resulting in higher power output (see Fig. \ref{fig:exp}a).

\subsection{Experimental realization}
\label{sec:exp}
Major advances in realizing microscopic thermal machines have been made in recent years. 
Experimental demonstration of such devices have been implemented with trapped ions \cite{rossnagel16,maslennikov17}, superconducting circuits \cite{pekola07,fornieri16}, and quantum dots \cite{thierschmann15}.
In this section we will concentrate on the pioneering experiment by Klatzow et al. \cite{klatzow17} in nitrogen-vacancy centers in diamonds, which clearly demonstrates the contrast between classical and quantum heat engines.
In particular, klatzow et al. have demonstrated both the equivalence of two-stroke and continuous engines and the violation of the stochastic power bound discussed in previous sections. 
\par
\begin{figure}
\center{\includegraphics[width=8cm]{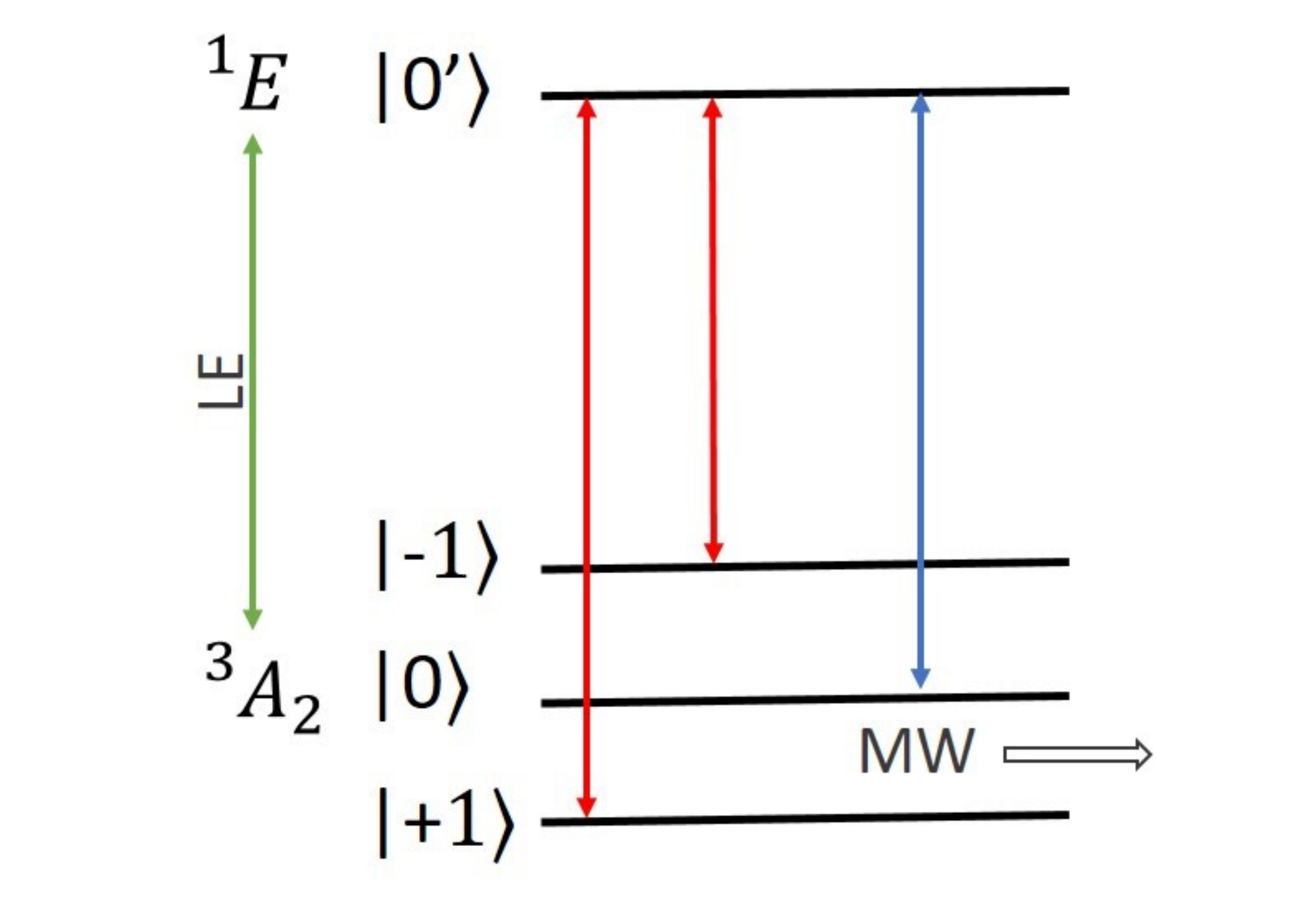}}
\caption{Scheme of NV$^-$ center levels and the effective couplings. The ground state triplet and the meta-stable exited singlet state are coupled through a 532 nm laser field. The collective effects of optical excitation, de-excitation and non-radiative decay are similar to coupling the NV$^-$ to  hot (red arrows) and cold (blue arrow) thermal baths. Work is extracted in the form of amplifying a MW field that is in resonance with the transition $\ket{+1} \leftrightarrow \ket{0}$.}
\label{fig:NV}
\end{figure}

The working medium of the engine is an ensemble of negatively charged nitrogen vacancy (NV$^-$) centers in diamond.  
The NV$^-$ center consist of a ground state spin triplet ${}^3A_2$ $\{\ket{-1}, \ket{0}, \ket{+1} \}$  in which degeneracy can be removed by applying a magnetic field.
The excited states consist of two spin triplets and three singlets.  
Due to fast decay rates and time averaging at room temperature, the excited states can be considered as a meta-stable spin singlet state ${}^1E$ denoted $\ket{0'}$.
Optical excitation, de-excitation (fluorescence) and non-radiative decay through the state $\ket{0'}$ mimic the effect of the couplings to the thermal baths.
The system approaches a steady state with population inversion between the states $\ket{0}$ and $\ket{+1}$.
Schematic description of the system is introduced in Fig. \ref{fig:NV}.

\par
A microwave (MW) field  set in resonance with the  $\ket{+1} \leftrightarrow \ket{0}$ transition is then amplified, and power in the form of stimulated emission of MW radiation (maser) is extracted from the engine.
The power is determined from a direct measurement of the fluorescence intensity change once the MW field is applied.
This setup can be mapped to the three-level continuous engine discussed in detail in Sec. \ref{sec:heat_mach_type}. 
Since the level $\ket{-1}$ is off-resonance with the MW field, it will not contribute to work extraction. 
Then the states $\{ \ket{+1}, \ket{0}, \ket{0'} \}$ in Fig. (\ref{fig:NV}) can be mapped to states $\{ \ket{1}, \ket{2}, \ket{3} \}$ of Fig. (\ref{fig1}), where the level $\ket{-1}$  only contributes to population transfer through level $\ket{0'}$.
Switching on and off the MW field allows for the realization of a two-stroke engine and its comparison to the continuous engine.
\par
\begin{figure}
\center{\includegraphics[width=14cm]{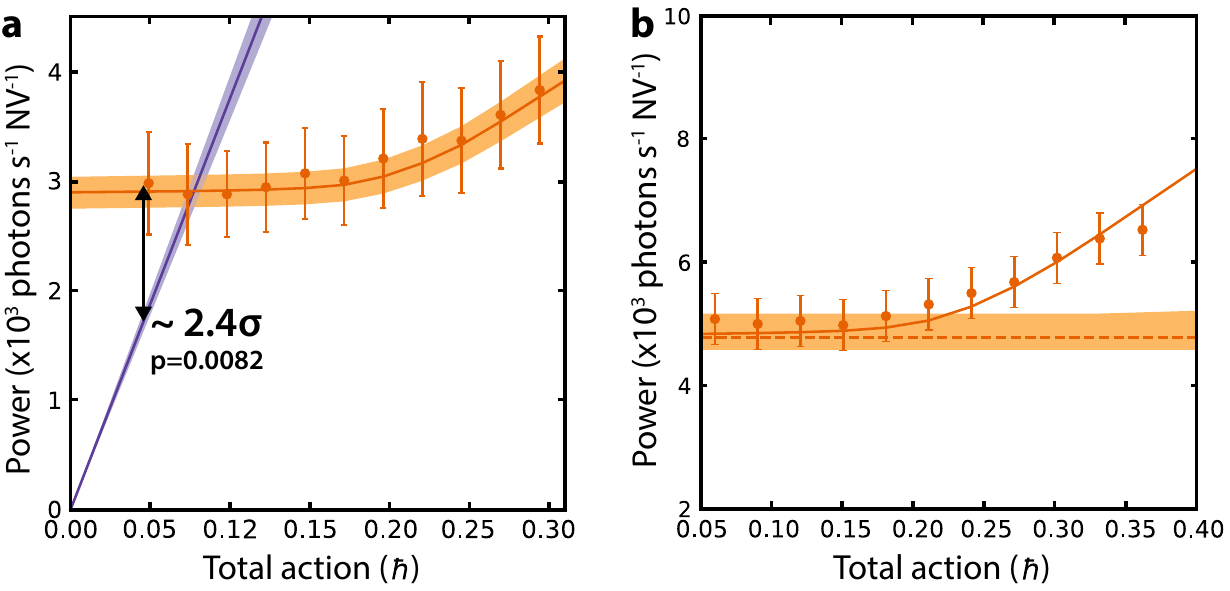}}
\caption{Power per NV$^-$ center vs. the normalized action $\bs$. (a) Quantum-thermodynamic signature: the orange line and data points present the theoretical and  measured power of the two-stroke engine, respectively. The purple straight line is the upper stochastic bound on the power. In the small action regime, a clear violation of the stochastic bound (2.4 standard deviation) is observed.   (b) The equivalence principle: The data points and the shaded region present power measurements of the two-stroke and continuous engines, respectively. The theoretical predictions for the two-stroke and continuous engines are given by the solid and dashed line, respectively.
In the small action regime, the two measurements coincide. Figure courtesy of  J. Klatzow, J. N. Becker, and E. Poem.}
\label{fig:exp}
\end{figure}

   
The normalized action in this case  \cite{klatzow17} is given by $\bs = [\tfrac{\epsilon}{2}d + \gamma (1-d)]\tau_{cyc}$, where $\epsilon$ in Eq. (\ref{eq:hamiltonian}) is twice the Rabi frequency, $\gamma$ is the total coupling rates to the baths and $d$ is the fraction of time of the work stroke out of the full cycle.
Fig \ref{fig:exp}a presents a power measurement of a two-stroke engine that exceeds the stochastic bound set by Eq. (\ref{eq:up_bound2}). 
This is a clear indication of a quantum-thermodynamic signature in the operation of the engine. 
The stochastic bound, plotted in the purple line,  was calculated for the  parameters $\omega=2\pi\times 2600$ MHz, $\epsilon=3.2$ Mrad/s, $\gamma=0.41$ MHz, and $d=1/3$ while changing the cycle time $\tau_{cyc}$. The bound is violated by $2.4$ standard deviations.
Fig. \ref{fig:exp}b demonstrates the equivalence principle in the form of Eq.(\ref{eq:power_equiv}). 
The powers of the two-stroke and continuous engines coincide in the small action regime. We also notice that, as  predicted, the power of the continuous engine is independent of the cycle time. 
That is, for $\tau_{cyc}\rightarrow 0$ the power does not vanish as it would for a stochastic engine in the small action regime. This indicates that a continuous engine with the Hamiltonian structure (\ref{eq:Hw_struc}) requires coherence in order to operate. 
In this experiment $\epsilon=5$ Mrad/s.
\par
This experiment demonstrated  the existence of quantum phenomena in microscopic thermal machines for the first time. 
In the small action regime, quantum coherence plays an important role in the operation of the engine. 
In this regime, the extracted power is enhanced, and thermodynamic properties of different types of engines coincide.

\subsection{Quantum friction}
Quantum friction is a quantum phenomenon that influences the operation of quantum thermal machines.
 It is related to the generation of coherences  and excitations along a finite-time unitary evolution, and its dissipation when coupled to a thermal environment.
Quantum friction in the context of quantum-theromodynamics was termed by Feldmann and Kosloff \cite{k152,k176} and is often referred as intrinsic or internal quantum friction.
To understand  this phenomenon intuitively, we will consider a specific example. 
We will compare two thermodynamic processes that will reveal the amount of energy lost to friction. 
The first process is an ideal quantum adiabatic process, where the initial state of the system is a thermal state $\rho_i = \exp[-\beta_i H_i]Z_i^{-1}$ with the initial inverse temperature $\beta_i$ and the initial Hamiltonian $H_i=H(0)$.
The system is then driven adiabatically by changing the system energy scale and keeping its population constant. 
We  require the final state to be expressed as $\rho_f = \exp[-\beta_f H_f]Z_f^{-1}$ such that $\beta_f$ is the inverse final temperature, $H_f=H(t_f)$, and the final populations satisfy $P_n^i=P_n^f$. 
This requirement implies that for all $n$-levels, the compression ratio is given by $\epsilon_n^f/\epsilon_n^i=\beta_i / \beta_f$, where $\epsilon_n^{f,i}$ are the eigenvalues of $H_{f,i}$. 
This condition corresponds to Sec. \ref{sec:energy_quant} homogeneous energy scaling.
\par
In real, finite-time processes, the adiabatic limit does not hold in general. 
Because of the non-commutativity of the Hamiltonian at different times, $[H(t),H(t')]\neq 0$, coherences and excitations will be generated  in the final state.
Thus, we will consider a second, more real process. In the first stage the system is driven non-adiabaticaly by changing the Hamiltonian $H_i\rightarrow H_f$ so that at the end of this stage the system is found in some state $\rho_{\tau}$.
In the second stage, the system is brought to equilibrium via coupling to a thermal bath at inverse temperature $\beta_f$ such that the final state is $\rho_f = \exp[-\beta_f H_f]Z_f^{-1}$.
Such a process is elementary when studying quantum thermal machines, such as in the study of Carnot and Otto cycles.
The energy that was invested in the creation of coherences and excitations and that was later dissipated to the thermal bath is the amount of energy lost to quantum friction.
\par   
In this sense, the effect of the non-adiabatic driving resembles the effect of friction in classical mechanics. 
Extra energy is needed to complete the process, which is then dissipated to the environment.  
The energy lost to friction is evaluated by the difference between the work invested in the real non-adiabatic stage and the work of the ideal adiabatic process,
\beq
W_{fric}=W_{re}-W_{ad}.
\eeq 
 This amount of work is exactly the amount of energy dissipated (irreversibly)  to the bath during the second stage of the second process.
 The work in the adiabatic process is $W_{ad}=tr(\rho_f H_f)-tr(\rho_i H_i)=\sum_n P_n^i(\epsilon_n^f-\epsilon_n^i)$.
For the non-adiabatic stage in the second process we have, $W_{re}=tr(\rho_{\tau}H_f)-tr(\rho_i H_i)$, implying that 
\beq
\label{eq:w_fric}
W_{fric}=\sum_n \epsilon_n^f \left(\bra{\epsilon_n^f}\rho_{\tau}\ket{\epsilon_n^f} -P_n^f \right).
\eeq  
The work (\ref{eq:w_fric}) can be related to the relative entropy of the final state  $\rho_{\tau}$ at the end of stage one with the final thermal state $\rho_f$ at the end of stage two  \cite{zambrini14},
\beq
\label{eq:w_fric2}
W_{fric}=\beta_f^{-1}S(\rho_{\tau}\parallel\rho_f).
\eeq
To show this we, consider the the definition of the relative entropy 
\beqar
S(\rho_{\tau}\parallel\rho_f)& = &tr(\rho_{\tau}\ln \rho_{\tau})-tr(\rho_{\tau}\ln\rho_{f}) = \sum_n P_n^i \ln P_n^i - \left(\bra{\epsilon_n^f}\rho_{\tau}\ket{\epsilon_n^f} - \ln P_n^f \right) \\ \nonumber
& = & \sum_n \beta_f \epsilon_n^f \left(\bra{\epsilon_n^f}\rho_{\tau}\ket{\epsilon_n^f} -P_n^f   \right).
\eeqar
Here we used the fact that the von-Neumann entropy is invariant to unitary transformations, the relation $P_n^i=P_n^f$, and that $\ln P_n^f = -\beta_f \epsilon_n^f-\ln Z_f$.
Since the relative entropy is non-negative,  Eq. \ref{eq:w_fric2} implies that $W_{fric} \geq 0$.
Applying arguments from \cite{deffner10} it was shown in \cite{zambrini14} that a tighter bound for $W_{fric}$ can be obtained using the Bures length
\beq
W_{fric}\geq \frac{8\beta_f^{-1}}{\pi}L^2(\rho_{\tau},\rho_f),
\eeq  
where $L(\rho_{\tau},\rho_f)= \text{arcos}(\sqrt{F(\rho_{\tau},\rho_f)})$, with the fidelity $ F(\rho_{\tau},\rho_f) = \left[tr(\sqrt{\sqrt{\rho_{\tau}}\rho_f \sqrt{\rho_{\tau}}})\right]^2$.
\par
More generally (i.e., not limited to the process above), the power along the work stroke of a thermal machine can be divided into  classical and coherent extraction mechanisms \cite{seifert17}
\beq
P \equiv tr\left(\frac{dH(t)}{dt}\rho(t) \right) = \sum_n \dot{\epsilon}_n(t)\bra{\epsilon_n(t)}\rho(t)\ket{\epsilon_n(t)}+\bra{\epsilon_n(t)}[H(t),\rho(t)]\ket{\dot{\epsilon}_n(t)}. 
\eeq   
 The first term can be associated with classical power,
\beq
\label{eq:clas_p}
P_{clas}=\sum_n \dot{\epsilon}_n(t)\bra{\epsilon_n(t)}\rho(t)\ket{\epsilon_n(t)}=\sum_n \dot{\epsilon}_n(t) \rho_{nn}(t),
\eeq
and the second term with the coherent power,
\beq
\label{eq:coh_p}
P_{coh}=\sum_n \bra{\epsilon_n(t)}[H(t),\rho(t)]\ket{\dot{\epsilon}_n(t)} = \sum_{n,l}(\epsilon_n-\epsilon_l)\braket{\epsilon_l}{\dot{\epsilon}_n}\rho_{nl}(t),
\eeq
where $\rho$ is expressed in the instantaneous eigenbasis of $H(t)$. 
Here, the term classical power is referred to power  generated by changes in the energy levels of the engine and the contribution comes only from the diagonal terms of the density matrix in the energy basis.
The contribution to the coherent power, however, arises only from the off-diagonal terms. 
The average classical and coherent work can be calculated by taking the time integral over the power.
For the adiabatic process described above, $P_{coh}$ vanishes and the population $\rho_{nn}$ is fixed in time. Thus, the average work is again given by $\sum_n P_n^i(\epsilon_n^f-\epsilon_n^i)$.
Starting with a diagonal state in the energy basis, which often occurs after a full thermalization stroke, coherence will be generated once $[H(t),H(t')]\neq 0$. Thus, $P_{coh} \neq 0$, and in general, the average coherent work will be non-zero.
The terms ``classical'' and ``coherent'' in Eq. (\ref{eq:clas_p}) and Eq. (\ref{eq:coh_p}) should be  used carefully.
As we saw in the previous section, coherences and populations are coupled to each other during the dynamical evolution, so the distinction between the two may become ambiguous.
\par
As long as dissipation is not introduced, friction is absent from the process, since the dynamic is reversible. 
In principle, one can extract all of the work from the state before it interacts with the thermal bath. 
By introducing a unitary cyclic process that transforms the state to a passive state 
\footnote{A passive state $\rho_{pas}$ is a state for which no work can be extracted  by a cyclic unitary operation, thus $tr(\rho_{pas}H) \leq tr(U\rho_{pas}U^{\dagger}H)$. 
It was shown by Pusz and Woronowicz that a state is passive if and only if it is diagonal in the energy eigenbasis and its eigenvalues are non-increasing with energy.} 
\cite{pusz78}, work can be extracted without changing the energy levels.
This is the maximal amount of extractable work from a quantum state, and is sometimes referred to as ergotropy \cite{allahverdyan04}.   
This procedure assumes that we have access to a semi-classical field and that the unitary operation can be implemented in practice.
%
\par    
Extracting work from coherences before they dissipate is one way to avoid friction.
Other strategies to avoid friction in finite-time processes employ different control methods. 
Quantum lubrication \cite{k215} is a method that minimizes $S(\rho_{\tau}\parallel\rho_f)$ in Eq. (\ref{eq:w_fric2}) by adding pure dephasing noise to the unitary stage. 
This additional noise can be considered as an instantaneous quantum Zeno effect, where the state of the system is projected to the instantaneous eigenstates of the time dependent Hamiltonian, eliminating the generation of coherence \cite{levy17}.       
Shortcut to adiabaticity \cite{torrontegui13} are other useful control methods that can be essisted in order to find frictionless control Hamiltonians \cite{muga09,delcampo14}, which speed up adiabatic processes. 
For a given final time, the state of the system is an eigenstate of the final Hamiltonian $H_f$, but during the process, the state is subject to the generation of coherences and excitations. 
These control processes can be executed on very short time scales with high fidelity.
If one assumes that the unitary evolution is not subject to noise from the control fields or from the environment, an exact frictionless Hamiltonian for finite time can be found. 
In a more realistic scenario  where noise is  present in the unitary process, one should aim to find a control Hamiltonian that minimizes the effect of noise \cite{levy18}.
\par
The average work measured by the energy difference between the end and the beginning of the process is the same for the shortcut to adiabticity protocol and the regular adiabatic process, which implies that frictionless protocols can in principle be implemented.
However, in the shortcut protocol the instantaneous power can become very large and will require different resources in order to be implemented.
Very fast protocols may require infinite resources.
This is the manifestation of the time-power trade-off  \cite{levy18} or the speed-cost trade-off \cite{deffner17}. 
This feature was also recently studied by examining the growth of work fluctuations in shortcut protocols \cite{del_campo17}.

\section{Quantum correlations in the operation of thermal machines}
The existence of quantum correlations is one of the most powerful signatures of non-classicality in quantum systems, as seen in recent reviews  \cite{modi12,horodecki09} and references therein.
In this context, thermodynamic approaches play a major role in witnessing and quantifying the existence of quantum correlations \cite{oppenheim02,zurek03,jennings10,terno10,popescu97,goold16}.  
Many studies relating thermodynamics with quantum correlation have focused on the amount of work extractable from correlations \cite{alicki13,karen13,marti15,campbell15} or the thermodynamic cost of creating correlations \cite{Huber15}.
\par
Quantum correlations have also been considered in the study of quantum thermal machines such as engines and refrigerators.
In \cite{brunner14}, entanglement between three qubits that form a quantum absorption refrigerator was studied.  
It was shown that near the Carnot bound entanglement is absent, nevertheless, in other regimes entanglement enhanced cooling and energy transport in the device. 
On the other hand, in a similar setup, that employed a different dynamical description of the refrigerator \cite{palao13}, it was shown that bipartite entanglement is absent between the qubits and correlations in the form of quantum discord which always exist in the system, do not influence the stationary heat flows. 
Entanglement in the working medium of quantum engines has mainly been investigated in the form of thermal-entanglement of two spins in Heisenberg types of models \cite{zhang07,ozgur14,wang09,huang12}.
In these studies, the heat and the work produced by the engines are expressed in terms of the concurrence that measure the amount of entanglement in the system.
Existence of temporal quantum correlations in a single two-level system Otto engine was also considered \cite{lutz17}. 
It was shown that in some regimes of the engine operations, the Legget-Garg function exceeds its maximal classical value, which is a  signature of non-classical behavior.
\par    
However, in most of the  studies referred above, quantum correlations do not play a major role in the thermodynamic operation of the devices. 
The correlations were not exploited to gain power, and no clear quantum-thermodynamic signature is obtained. 
Instead, these studies only indicates that quantum correlation are present in the working medium when setting carefully the parameters of the problem.     
Next, we review a quantum-thermodynamic signature in the form of heat exchange. 
We believe this work will motivate further study on whether and how correlation can be exploited in the repeated operation of quantum thermal machines.   

\subsection{Quantum-thermodynamic signatures}
Here we derive a result obtained by Jennings and Rudolph \cite{jennings10} on the average heat exchange between two local correlated thermal states. 
This study was motivated by the work of Parovi on the disappearance of the thermodynamic arrow in highly correlated environments \cite{partovi08}.
Although the setup is not a standard\footnote{Here we consider a ``standard'' thermal machine to be a device that has specific functioning, such as heat engines and refrigerators, and that operates in the limit cycle/steady state.}
thermal machine, which is the focus of this chapter, it can be considered as a single-shot cooling process where no external work is invested. 
Instead, correlations are exploited (as fuel) in order to cool the system. 
Moreover,  this study presents an important thermodynamic signature of quantum-mechanical features that goes beyond what is possible classically. 
In the language of quantum information, the signature is called an entanglement witness. 
In particular, a bound on the amount of heat that can flow from a cold local thermal state to a hot local thermal state due to classical correlations is obtained. 
It is further shown that quantum correlations may violate this bound.
\par
The Clausius statement of the second law of thermodynamics implies that when no work is done on the system, heat can only flow from a hotter body to colder body. 
This statement holds when no initial correlations are present between the two bodies.
To show this, we consider two initially uncorrelated systems $A$ and  $B$. 
The global initial state is thus represented by the product of the states of the  two subsystems, $\rho_{AB}(0)=\rho_{A}(0)\otimes \rho_{B}(0)$. This is known as Boltzmann's assumption of molecular chaos.
We further assume that each subsystem is initially in a thermal state, $\rho_X(0) = \exp[-\beta_X H_X]$, where $X \in \{A,B\}$.
Next we let the two subsystems interact via a global unitary interaction $U(\tau,0)$ for time $\tau$ that couples the two subsystems. 
As we wish to preserve the average energy of $\rho_{AB}$, and thus not perform any work on the system from external sources, we require that the transformation uphold  $tr\left[ \rho_{AB}(0)(H_A+H_B)\right]=tr\left[ \rho_{AB}(\tau)(H_A+H_B)\right]$.
After the interaction, the local reduced states are $\rho_{A,B}=tr_{B,A}(\rho_{AB})$ and the corresponding internal energy and entropy will change by the amount $\Delta E_X$ and $\Delta S_X$, where  
\beqar
\Delta E_X &=& tr\left[(\rho_X(\tau)-\rho_X(0))H_X \right],\\ \nonumber
\Delta S_X &=& S_X(\tau) -S_X(0), 
\eeqar
and  $S_X = -tr(\rho_X \ln \rho_X)$ is the von-Neumann entropy.
Since the thermal state minimizes the free energy, $F(\rho_X)=tr\left[ \rho_X H_X\right]-S(\rho_X)/ \beta_X$, we obtain, $tr\left[ \rho_X(0) H_X\right]-S(\rho_X(0))/ \beta_X \leq tr\left[ \rho_X(\tau) H_X\right]-S(\rho_X(\tau))/ \beta_X $, which can be cast into\footnote{Note that the von-Neumann entropy for nonequilibrium states cannot be associated with the thermodynamic entropy. 
However, the inequality can be derived directly from the relative entropy $S(\rho(\tau)\parallel\rho(0))$, where $\rho(0)$ is a thermal state.} \cite{peres06}
\beq
\label{eq:delta_f}
\beta_X \Delta E_X -\Delta S_X \geq 0.
\eeq
\par
To quantify the amount of correlations present between the two subsystems $A$
and $B$ we will use the quantum mutual information for bipartite systems:
\beq
I_q[\rho_{AB}]\equiv S(\rho_{AB} \parallel \rho_A \otimes \rho_B)=S_A +S_B-S_{AB}.
\eeq 
This function is always non-negative \cite{wehrl78} and accounts for both classical and quantum correlations.
For product states, like the one considered in the process above, the mutual information vanishes, $I_q[\rho_{AB}(0)]=0$. 
This implies that at time $\tau$ the change in the mutual information is $\Delta I_q[\rho_{AB}]=I_q [\rho_{AB}(\tau)] \geq 0$. 
On the other hand since $S_{AB}$ is invariant under the unitary transformation $U(\tau,0)$ we obtain $\Delta I_q[\rho_{AB}]=\Delta S_A+\Delta S_B$ which is true irrespective of the initial conditions. 
Applying these relation to (\ref{eq:delta_f}) we obtain
\beq
\label{eq:clasuis}
\beta_A Q_A + \beta_B Q_B \geq \Delta I_q(\rho_{AB}) \geq 0,
\eeq
where we identified the heat transferred to system $X$ as the change in its internal energy, $Q_X=\Delta E_X$. 
Since $Q_A=-Q_B$, relation (\ref{eq:clasuis}) implies that heat will always flow from hot to cold. 
Note, however, if the assumption of initial product state is removed then $\Delta I_q[\rho_{AB}]$ can become negative, which opens up the possibility of ``anomalous heat flow'' namely heat flowing from the colder body to the hotter body. 
In this case Eq. (\ref{eq:clasuis}) reads
\beq
\beta_A Q_A + \beta_B Q_B \geq -|\Delta I_q(\rho_{AB})|.
\eeq
\par
The initial correlations present in the system, which can be either classical, quantum, or both, are the source of this backward flow. 
In order to derive a clear quantum-thermodynamic signature, a bound on the maximal backward flow that could occur from just classical correlations is set.
The classical mutual information of a quantum state is given by taking the maximum over all POVMs $M_A\otimes M_B$, which are the most general kind of quantum measurements possible on $A$ and $B$   \cite{horodecki02}:   
\beq
I_c(\rho_{AB})= \max_{M_A\otimes M_B} \left[ H(P_A)+ H(P_B)-H(P_{AB})\right],
\eeq
where $H(P)$ is the Shannon entropy of the measurement statistics $P$.
The classical mutual information satisfies\footnote{The relative entropy is monotonically decreasing under completly positive maps.  } $0 \leq I_c(\rho_{AB}) \leq I_q(\rho_{AB})$ \cite{Lindblad74}, which implies that $\max |\Delta I_c(\rho_{AB})| \leq \max |\Delta I_q(\rho_{AB})|$. 
To obtain a bound on $\max |\Delta I_c(\rho_{AB})|$ that is independent of the specific details of the states, we note that $I_c(\rho_{AB}) \leq \ln D$, where $D =\min\{ \text{Dim}(\rho_A),\text{Dim}(\rho_B)\}$ is the dimension of the smaller system.
This bound is a result of the monotonicity property of the Shannon entropy under partial trace, $H(P_{AB})\geq H(P_{A}), H(P_{B})$.
The same argument applies for the von-Neumann entropy for separable states $S(\rho_{AB}|_{\text{sep}})\geq S(\rho_{A}), S(\rho_{B})$ \cite{HORODECKI94}.
The upper bound, $\ln D$, is saturated for perfectly correlated, zero discord, separable states $\rho_{AB}=\tfrac{1}{D}\sum_k \ketbra{e_k}{e_k}_A\ketbra{f_k}{f_k}_B$, where $\ket{e_k}$ and $\ket{f_k}$  are orthonormal bases for A and B. 
Assuming that the initial state $\rho_{AB}$ is the perfectly correlated state above and that at the final time it is a product state,  the maximal violation of heat transfer from cold to hot due to classical correlations is
\beq
Q_{clas}=\frac{\ln D}{ |\beta_A -\beta_B|}.
\eeq   
Since for entangled states the mutual entropy can exceed $\ln D$, and in fact, go to $2 \ln D$ for maximally entangled states, the violation of heat transfer can exceed $Q_{clas}$. 
Thus, any measurement  of heat exchange that gives $Q > Q_{clas}$ indicates that the initial state $\rho_{AB}$ was necessarily entangled. 
Therefore the heat-flow pattern acts as a ``witness'' that reveals genuine non-classicality in the thermodynamic system.  Moreover, this quantum resource can be exploited in a cooling process, removing heat from a colder body and transfer it to a hotter one.           
This phenomenon was recently demonstrated in a NMR experiment \cite{micadei17}, where it was shown that heat is transferred from a colder nuclear spin to a hotter one due to correlations. 
The result presented here was later extended to derive an exchange fluctuation theorem for energy exchange between thermal quantum
systems beyond the assumption of molecular chaos \cite{jennings12}.
\section{Conclusions}
In this chapter we have presented a short review about some of the studies
that compare the performance of classical and quantum heat machines. 
In particular, we have focused on three main quantum features, energy quantization, quantum coherence, quantum correlations, and their manifestation in the operation of quantum thermal machines. 
All of these features boil down to the mathematical structure of quantum mechanics, which is represented by wave functions and operators acting in Hilbert space.  However, we showed that  the manifestation of particular quantum phenomena  can be distinguished  from one another.
Specifically, we saw that energy quantization and the uncertainty principle alone lead to differences in the behavior of heat engines. 
Moreover, quantum mechanics enables the realization of thermal-machines with incompressible working medium which is classically impossible. 
\par
The role of quantum coherence was also studied. 
Since the generation of coherences for interacting quantum systems is typically unavoidable, and since coherences have thermodynamic cost, dephasing leads to quantum friction.
Working in the small action regime, complete dephasing can be suppressed. 
In this regime, coherences survive the couplings to the thermal baths, and  quantum engines outperform stochastic engines, which incorporate only populations in their operation. This phenomenon can be observed experimentally by setting a bound on the maximal power obtained by a stochastic engine. 
A violation of this bound is a signature of a quantum-thermodynamic behavior.
Furthermore, in the small action regime, the thermodynamic properties of different types of engines coincide, since coherent work extraction becomes dominant.   
\par
Quantum correlations are obviously closely related to the existence of coherence between particles, however, by investigating bounds on classically correlated systems, it is possible to isolate the effects of quantum correlations. Here, we demonstrated that the presence of quantum correlations can explain anomalous heat flow that cannot be achieved classically.     
\par
As we have shown, operating in the quantum regime is not always advantageous.
Thus it is important to understand the quantum behavior of thermal-machines in order to avoid undesired reductions in the performance. 
We believe that the results presented here represent only some initial results that require further investigation, such as: 
\begin{itemize}
\item What are the features needed on the potentials for providing a quantum
boost to the work and heat exchange and how do these features relate
to other quantum proprieties? 
\item One possible manner of obtaining an efficiency divergence between classical and quantum heat machines is by performing a thermal cycle that involves an inhomogeneous energy level scaling. 
As of today, there are only a small number of studies concerned with this type of cycle, calling for further research in this direction. 
\item Since coupling to a thermal source is an elementary process in thermal devices, innovative methods to exploit quantum coherences and correlations before they decay are essential for obtaining quantum supremacy. Possible approaches are:  identifying and operating in the regime where the quantum effects are still relevant; applying quantum control methods, such as feedback control, in order to maintain the quantum properties. 
\item The main results concerning the relationship between thermodynamics and quantum correlations are focused on single-shot processes. A desired goal would be to exploit correlations in the repeated operation of a device.
\item Can the quantum thermal-machines be scaled up in size and maintain their quantum nature? Can we deduce from these models something about energy transfer and quantum behavior of more complex systems, such as biological systems?      
\end{itemize}
\par
 As we have shown, the actual performances of classical and quantum heat machine diverge, even though both are limited by the same fundamental thermodynamic bounds, such as the Carnot efficiency. Nevertheless, the difference between them is noteworthy and should be studied further, especially considering the fast experimental progress that has already succeeded in demonstrating quantum behavior of thermal-machines.
\acknowledgements

We thank R. Kosloff, R. Uzdin, D. Jennings, M. lostaglio, D. Jasrasaria, A. Bylinskii, D. Gangloff, R. Islam, A. Aspuru-Guzik and V. Vuletic for useful comments and for sharing their wisdom. 
We also acknowledge J. Klatzow, J. N. Becker and E. Poem. for sharing their experimental results and preparing Fig. \ref{fig:exp}.

%


\end{document}